\documentclass{ieeetj}
\usepackage{cite}
\usepackage{amsmath,amssymb,amsfonts}
\usepackage{algorithmic}
\usepackage{graphicx,color}
\usepackage{textcomp}
\usepackage{xcolor}
\usepackage{hyperref}
\hypersetup{hidelinks=true}
\usepackage{algorithm,algorithmic}
\def\BibTeX{{\rm B\kern-.05em{\sc i\kern-.025em b}\kern-.08em
    T\kern-.1667em\lower.7ex\hbox{E}\kern-.125emX}}
\AtBeginDocument{\definecolor{tmlcncolor}{cmyk}{0.93,0.59,0.15,0.02}\definecolor{NavyBlue}{RGB}{0,86,125}}

\usepackage[caption=false,font=footnotesize]{subfig}

\usepackage{stfloats}

\usepackage{placeins}
\usepackage{float}

\usepackage{tabularx,colortbl,multirow}
\usepackage{enumerate}

\def\OJlogo{\vspace{-4pt}$<$Society logo(s) and publication title will appear here.$>$}
\def\seclogo{\vspace{10pt}$<$Society logo(s) and publication title will appear here.$>$}

\def\authorrefmark#1{\ensuremath{^{\textbf{#1}}}}

\begin{document}
\receiveddate{XX Month, XXXX}
\reviseddate{XX Month, XXXX}
\accepteddate{XX Month, XXXX}
\publisheddate{XX Month, XXXX}
\currentdate{XX Month, XXXX}
\doiinfo{XXXX.2022.1234567}

\markboth{}{Author {et al.}}

\title{Physics-Informed Deep Recurrent Back-Projection Network for Tunnel Propagation Modeling}

\author{Kunyu~Wu\authorrefmark{1}, Qiushi~Zhao\authorrefmark{1}, Jingyi~Zhou\authorrefmark{1}, Junqiao~Wang\authorrefmark{1}, Hao~Qin\authorrefmark{1, 2}, Xinyue~Zhang\authorrefmark{2}, \\ and Xingqi~Zhang\authorrefmark{3}, Senior~Member, IEEE}
\affil{School of Electronics and Information Engineering, Sichuan University, Chengdu, China.}
\affil{School of Electrical and Electronic Engineering, University College Dublin, Dublin, Ireland.}
\affil{Department of Electrical and Computer Engineering, University of Alberta, Canada T6G 2H5.}
\corresp{Corresponding author: Hao Qin (email: hao.qin@scu.edu.cn).}
\authornote{This paragraph of the first footnote will contain support information, including sponsor and financial support acknowledgment.}


\begin{abstract}
Accurate and efficient modeling of radio wave propagation in railway tunnels is is critical for ensuring reliable communication-based train control (CBTC) systems. Fine-grid parabolic wave equation (PWE) solvers provide high-fidelity field predictions but are computationally expensive for large-scale tunnels, whereas coarse-grid models lose essential modal and geometric details. To address this challenge, we propose a physics-informed recurrent back-projection propagation network (PRBPN) that reconstructs fine-resolution received-signal-strength (RSS) fields from coarse PWE slices. 
The network integrates multi-slice temporal fusion with an iterative projection/back-projection mechanism that enforces physical consistency and avoids any pre-upsampling stage, resulting in strong data efficiency and improved generalization.
Simulations across four tunnel cross-section geometries and four frequencies show that the proposed PRBPN closely tracks fine-mesh PWE references. 
Engineering-level validation on the Massif Central tunnel in France further confirms robustness in data-scarce scenarios, trained with only a few paired coarse/fine RSS.
These results indicate that the proposed PRBPN can substantially reduce reliance on computationally intensive fine-grid solvers while maintaining high-fidelity tunnel propagation predictions.
\end{abstract}




\maketitle

\section{INTRODUCTION}
\IEEEPARstart{R}{adio} wave propagation modeling in tunnel environments is crucial for the effective deployment of intelligent transportation systems (ITS), particularly for communication-based train control (CBTC) systems along railways \cite{7836324, 10005074, liu20246g, wu2025lessaccess, Qin25UAV, Guan13, 9284448, 6808529,6387578}. Deterministic propagation models based on parabolic wave equation (PWE) methods \cite{8487053, Qin23_AWPL_SSSPE, 7875498, OZGUN20112638, Qin23comparative, 11177968} have been widely utilized due to their high accuracy in handling long guiding structures such as tunnels. However, as wireless communication systems transition to higher frequency bands, the PWE methods necessitate smaller discretization steps, leading to significant computational costs and time, which pose challenges for real-time applications and economic feasibility \cite{Huang24Frequency, QIN25TOA, 11186207}. Therefore, techniques to significantly accelerate high-frequency propagation modeling have become increasingly critical and timely.

To overcome these challenges, machine learning (ML) techniques have shown considerable potential in modeling radio wave propagation within tunnel environments in recent years \cite{10144543, 9496115, LI2022268, 10066315, Aldossari2019, Bakirtzis22, 9713743, wang2022vector, 9528707}. An approach proposes a long-range attentive propagation network model informed by generative physics (PLAPN) \cite{11198828}, which integrates convolutional feature encoding, dual-path attention, and a sliding window inference strategy. This method achieves high-precision long-range received signal strength (RSS) predictions in tunnels using only the initial 10\% of PWE simulation results. Another approach introduces an inception-enhanced generative adversarial network (Inc-GAN) \cite{zhou2025physicsconst}, which reconstructs electric field distributions across tunnel cross-sections using sparse data from actual train operations, significantly improving computational efficiency compared to traditional methods. Furthermore, a physics-guided high-fidelity modeling framework has been proposed \cite{10858646}, capable of adapting to various terrain types and antenna configurations. By embedding prior knowledge from deterministic models into the network architecture, this framework addresses the computational intensity limitations of the split-step parabolic equation (SSPE) method for radio wave propagation in irregular terrains, significantly enhancing both predictive accuracy and generalization capability.

However, most ML-assisted tunnel propagation models require large, curated datasets that are expensive and difficult to acquire. Simulated data at mmWave with fine PWE discretization scales poorly with frequency and domain size, making bulk production infeasible under realistic compute and time budgets. Field measurements are limited by safety rules, scarce maintenance windows, and coordination overheads, hindering datasets that span diverse geometries, materials, frequencies, and kilometer ranges; domain shifts further erode generalizability. Meanwhile, architectures tuned for deployment latency are typically shallow with small kernels, restricting capacity to capture rich multipath and modal coupling, especially in curved or multi-branch tunnels. Capturing long-range context often requires deeper networks or heavy dilation, which sharply increases compute and complexity and induces optimization instability, larger memory footprints, and unpredictable latency.

To address these challenges, we propose a physics-informed deep recurrent back-projection propagation network (PRBPN), a physics-aware framework that elevates coarse-mesh PWE slices to fine-mesh RSS distributions by modeling the longitudinal wave evolution in a tunnel as a short matrices sequence. Unlike single-frame super-resolution, the model performs multi-frame-to-one prediction: a few preceding and succeeding coarse slices are fused to reconstruct the target fine slice, leveraging the inherent physical smoothness of wave propagation. The architecture integrates a compact encoder for coarse inputs, an iterative projection/back-projection refinement module that enforces LR-to-HR consistency via error feedback, and a lightweight reconstruction head for fine-mesh outputs. 
Thanks to these inductive biases and the exploitation of short-range temporal context, the model achieves high data efficiency: only a few paired coarse/fine samples suffice to produce high-fidelity predictions, even for large scale factors.


Building on the multi-frame prediction framework, PRBPN incorporates difference-weighted temporal fusion (DWTF) to aggregate information from neighboring coarse slices. The fusion leverages pixel-level differences to emphasize multipath-informative regions, such as reflections along tunnel walls, while suppressing misaligned or noisy cues from coarse simulations. The resulting representation is then refined through a physics-consistent error feedback mechanism: a provisional high-resolution estimate is back-projected to the coarse domain, its residual iteratively corrected, and the update propagated forward to stabilize learning and preserve energy conservation and propagation continuity. This coarse-to-fine correction captures long-range context without requiring deep or heavily dilated networks, improving reconstruction fidelity in long tunnel segments while maintaining tractable computation and memory footprints, making PRBPN suitable for real-time tunnel monitoring and deployment.
By leveraging the PRBPN architecture and DWTF, the method achieves high-fidelity reconstruction and outperforms both learning- and physics-based baselines. It remains applicable to long-range, data-limited, and dynamically evolving tunnel systems. Extensive quantitative metrics and visual comparisons corroborate these findings.



The main contributions are summarized as follows:
\begin{enumerate}
  \item We introduce PRBPN, which lifts coarse-mesh PWE slices to fine-mesh RSS distributions, fully exploiting the global RSS context to achieve accurate and efficient reconstructions with strong generalization across tunnel geometries and frequency bands.
  \item We develop a DWTF module coupled with a physics-consistent projection/back-projection refinement mechanism, enabling long-range context modeling while avoiding very deep or heavily dilated networks.
  \item We demonstrate high data efficiency, achieving competitive performance with as few as 10–20 coarse–fine RSS pairs.
  \item We provide validation using measurements from the Massif Central tunnel, confirming robustness and applicability in real-world deployment scenarios.
\end{enumerate}

The paper is organized as follows. Section II presents the proposed model and methodology. Section III provides validation across diverse tunnel scenarios. Section IV reports engineering validation using real tunnel measurements. Section V concludes the paper.

\section{Methodology and Model Description}\label{Section II}

\subsection{Overview of the Proposed Framework}

\begin{figure*}[ht]
    \centering
    \includegraphics[width=0.95\textwidth]{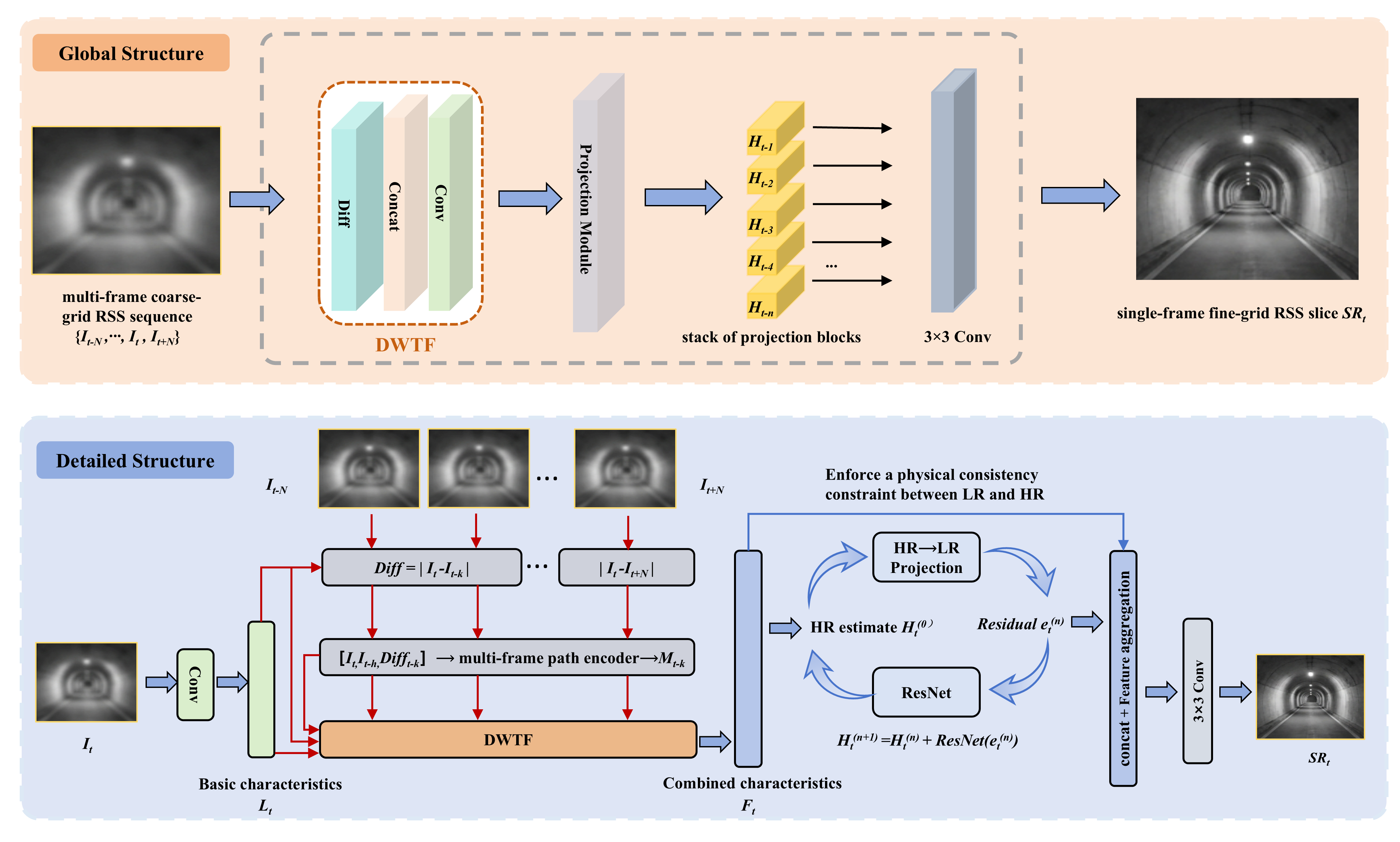}
    \caption{Overview of PRBPN architecture integrating single-slice and multi-slice paths with DWTF and projection/back-projection refinement for tunnel propagation.}
    \label{fig:architecture}
\end{figure*}

The proposed approach addresses the computational and adaptability limitations of traditional wave propagation models by integrating a lightweight architecture with a fully parallelized attention mechanism, specifically designed for tunnel-based RSS data. PRBPN treats the evolution of RSS as a short sequence, performing multi-frame-to-one prediction by fusing adjacent coarse-mesh slices to reconstruct the target fine-mesh slice, leveraging the physical smoothness of sequential propagation. The model avoids any upsampling pre-stages, directly using coarse-mesh PWE slices as inputs, with supervision derived from fine-mesh references or measurements. This design ensures strong data efficiency, requiring only 10–20 paired coarse/fine samples to achieve competitive accuracy, thanks to the back-projection inductive bias and short-range temporal context, which reduce reliance on large training datasets. The PRBPN architecture, shown in Fig.~\ref{fig:architecture}, highlights the encoder, attention module, and decoder components involved in the RSS prediction pipeline.


\subsection{Input Representation and Preprocessing}

PRBPN operates on sequences of coarse-mesh tunnel slices $\{I_{t-n}, \dots, I_t, \dots, I_{t+n}\}$, where $I_t$ denotes the target slice and each $I$ is a low-resolution RSS field produced by a coarse-mesh PWE discretization. PRBPN directly consumes the raw coarse-mesh data to avoid interpolation artifacts and better reflect practical constraints in high-frequency propagation modeling. For temporal fusion, inter-slice differences are computed pixel-wise between $I_t$ and $I_{t\pm k}$, and no explicit motion estimation is required; difference-weighted temporal fusion provides implicit alignment. All inputs are normalized to $[0,1]$; training uses random rotations, flips, and crops to improve generalization. For a scale factor of $\times 8$, convolutional kernel, stride, and padding are set to 12, 8, and 2, respectively, enabling large upsampling while mitigating errors associated with preliminary interpolation.

\subsection{Model Architecture}

The model consists of initial feature extraction, a PRBPN encoder, a difference-weighted temporal fusion module, a PRBPN decoder with projection/back-projection refinement, and a reconstruction head. The target slice $I_t$ follows a single-image path while multi-slice context is integrated through a multi-image path; both are refined iteratively to produce the final fine-mesh estimate.

\subsubsection{PRBPN Encoder}

The PRBPN encoder extracts and refines features from coarse-mesh inputs while encoding physics-informed biases such as propagation smoothness and local continuity. For the target slice $I_t$, initial features are obtained via a $3 \times 3$ ConvBlock with stride 1 and PReLU activation:
\begin{equation}\label{eq:enc-init}
L_t = \text{ConvBlock}(I_t)
\end{equation}

For each neighboring frame $I_{t-k}$, pixel-level differences are computed as:
\begin{equation}\label{eq:enc-diff}
\text{diff}_{t-k} = |I_t - I_{t-k}|
\end{equation}

These differences are then concatenated with $I_t$ and $I_{t-k}$ to form a multi-channel input that is processed by another ConvBlock, yielding the neighbor features:
\begin{equation}\label{eq:enc-neigh}
M_{t-k} = \text{ConvBlock}(I_t \oplus I_{t-k} \oplus \text{diff}_{t-k})
\end{equation}
where $\oplus$ denotes the concatenation operation.

\subsubsection{Difference-Weighted Temporal Fusion (DWTF)}

The DWTF module aggregates adjacent slices using attention derived from pixel-level differences between the target slice and its neighbors. For each neighboring slice $I_{t-k}$, the pixel-wise absolute difference with the target slice $I_t$ is computed:
\begin{equation}\label{eq:dwtf-diff}
\text{diff}_{t-k} = |I_t - I_{t-k}|
\end{equation}

A single-channel heatmap is generated by performing channel-wise averaging on $\mathrm{diff}_{t-k}$:
\begin{equation}\label{eq:dwtf-mean}
\text{mean\_chan}(\text{diff}_{t-k}) = \frac{1}{C} \sum_{c=1}^{C} \mathrm{diff}_{t-k}[c],
\end{equation}
where $C$ is the number of channels. This heatmap is then passed through a $3 \times 3$ convolution and a sigmoid activation function to generate an attention map:
\begin{equation}\label{eq:dwtf-attn}
\mathrm{attn}_{t-k} = \sigma(\mathrm{Conv}(\text{mean\_chan}(\mathrm{diff}_{t-k})))
\end{equation}

The attention map is used to weight the differences, emphasizing informative regions and suppressing irrelevant or noisy ones. The weighted difference is calculated as:
\begin{equation}\label{eq:dwtf-wdiff}
\mathrm{wdiff}_{t-k} = \mathrm{attn}_{t-k} \odot \text{mean\_chan}(\text{diff}_{t-k}),
\end{equation}
where $\odot$ denotes the element-wise multiplication. The weighted differences from all neighboring slices are then aggregated to form a context-aware feature representation for the target slice. This representation is passed through the encoder's multi-slice path, which provides enhanced temporal context for accurate RSS prediction.

This process allows the model to focus on regions of interest, such as high-gradient reflection zones, while suppressing noise or misaligned information. The fused representation is then refined through physics-consistent error feedback, where a provisional HR estimate is back-projected to the coarse domain, its residual is iteratively corrected, and the updated HR estimate is fed forward. This end-to-end correction improves reconstruction fidelity over long distances and in geometrically complex tunnel segments, while maintaining computational and memory efficiency for real-time or near-real-time deployment.

\subsubsection{PRBPN Decoder}

The decoder implements recurrent back-projection to enforce LR-to-HR consistency. A provisional HR estimate $H_t$ is projected back to the coarse domain, compared with the input to form a residual, refined by ResBlocks, and fed forward to update $H_t$ iteratively. The process is defined as:
\begin{equation}\label{eq:dec-enc}
H_t = \text{NetE}(L_{t-1}, M_{t-1}; \theta_E)
\end{equation}
\begin{equation}\label{eq:dec-dec}
L_t = \text{NetD}(H_t; \theta_D)
\end{equation}
where $\text{NetE}$ and $\text{NetD}$ are the encoder and decoder functions, respectively, and $\theta_E$ and $\theta_D$ are their parameters.

In the back-projection loop, the error $e_t$ is calculated as:
\begin{equation}\label{eq:dec-err}
e_t = H_t - L_t
\end{equation}

This residual is refined and added back to $H_t$ in an iterative manner:
\begin{equation}\label{eq:dec-iter}
H_t^{(n+1)} = H_t^{(n)} + \text{NetRes}(e_t; \theta_{\text{res}})
\end{equation}

Finally, the fine-mesh output is produced by the reconstruction ConvBlock:
\begin{equation}\label{eq:dec-out}
SR_t = \text{ConvBlock}([H_t^{(n)}, H_{t-1}^{(n-1)}, \dots, H_{t-n}^{(n-n)}])
\end{equation}

\subsection{Model Interpretation and Rationale}

PRBPN is designed for tunnel propagation where inputs are naturally coarse due to computational constraints. Directly handling coarse-mesh PWE slices preserves native structures and avoids artifacts introduced by preliminary interpolation. Difference-weighted temporal fusion supplies adaptive, physically motivated context by emphasizing regions where slice-to-slice evolution carries informative multipath cues. The projection/back-projection mechanism imposes a consistency prior between the predicted HR field and the observed coarse field, using residual correction to recover missing details while respecting local continuity. Formally, for target $I_t$ and neighbors $\{I_{t\pm k}\}$,
\begin{equation}\label{eq:model-map}
H_t = f_{\mathrm{dec}}\!\left(f_{\mathrm{DWTF}}\!\left(f_{\mathrm{enc}}(I_t,\{I_{t\pm k}\})\right)\right),
\end{equation}
with iterative updates driven by residuals between the back-projected estimate and the coarse input to promote physics-consistent refinement over long sequences and complex tunnel geometries.

\subsection{Loss Function}

The PRBPN model is optimized using an $L_1$ loss between the predicted HR output $SR_t$ and ground truth $HR_t$, as follows:
\begin{equation}\label{eq:loss-rec}
\mathcal{L}_{\mathrm{rec}} = \lVert SR_t - HR_t \rVert_1
\end{equation}

To ensure physical smoothness in the predicted fields, an optional total variation (TV) or gradient penalty term can be added:
\begin{equation}\label{eq:loss-smooth}
\mathcal{L}_{\mathrm{smooth}} = \lVert \nabla SR_t \rVert_1
\end{equation}

The total loss function is then:
\begin{equation}\label{eq:loss-total}
\mathcal{L} = \mathcal{L}_{\mathrm{rec}} + \beta \, \mathcal{L}_{\mathrm{smooth}}
\end{equation}
where $\beta$ is a small scalar controlling the weight of the smoothness penalty.

The $L_1$ reconstruction loss ensures the model minimizes the difference between predicted and ground truth high-resolution fields. The addition of the gradient penalty term, $\mathcal{L}_{\mathrm{smooth}}$, enforces smoothness in the predicted field, which is crucial for modeling continuous wave propagation. It prevents abrupt transitions, ensuring the field remains physically plausible. The regularization factor $\beta$ balances reconstruction accuracy with smoothness. Without smoothness, the model may produce unrealistic discontinuities. These combined losses help the model generate both accurate and physically realistic high-resolution fields.

\begin{figure*}[ht]
\centering
\subfloat[]{\includegraphics[width=0.24\textwidth]{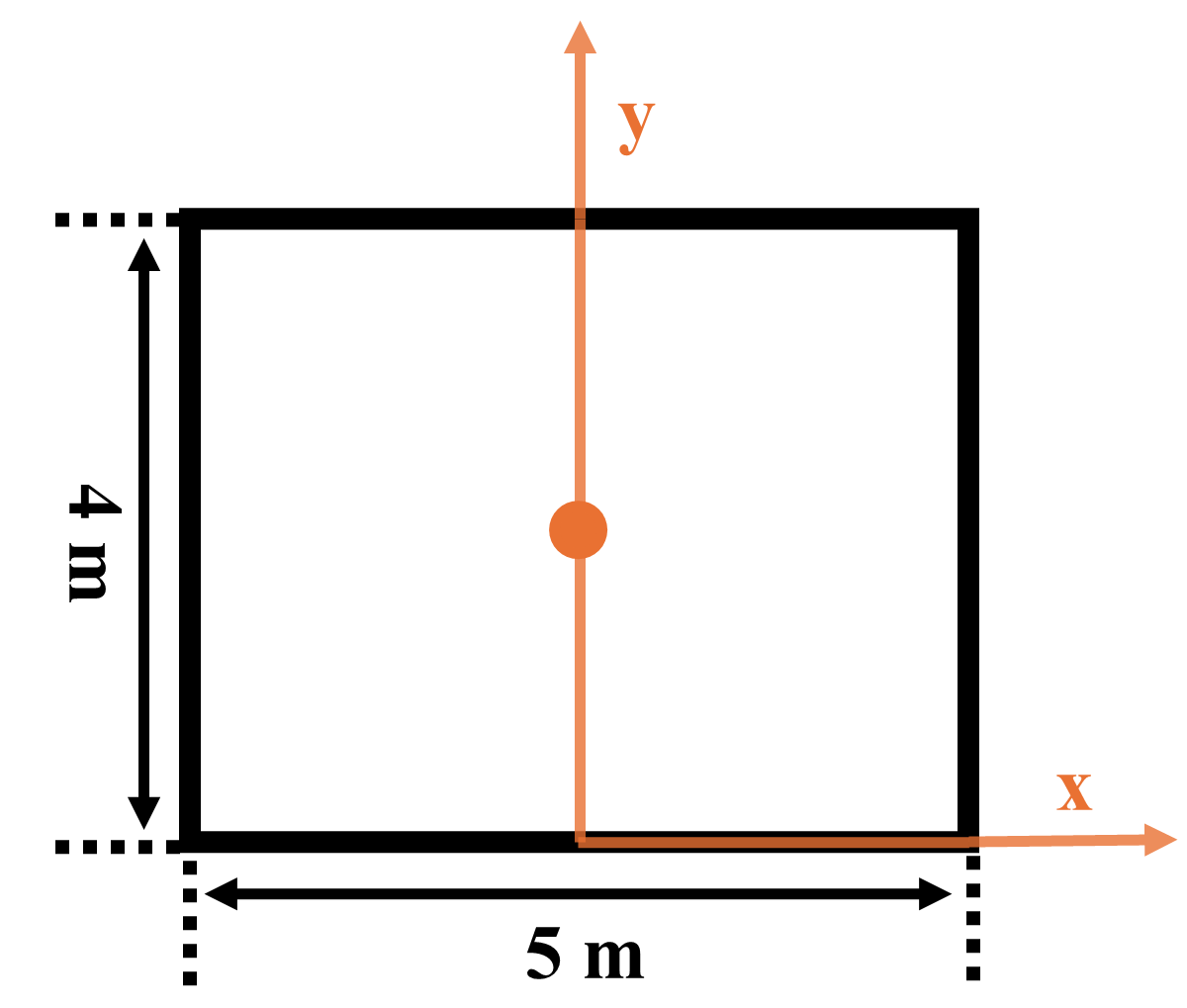}\label{fig:tunnelshape_a}}
\hfill
\subfloat[]{\includegraphics[width=0.24\textwidth]{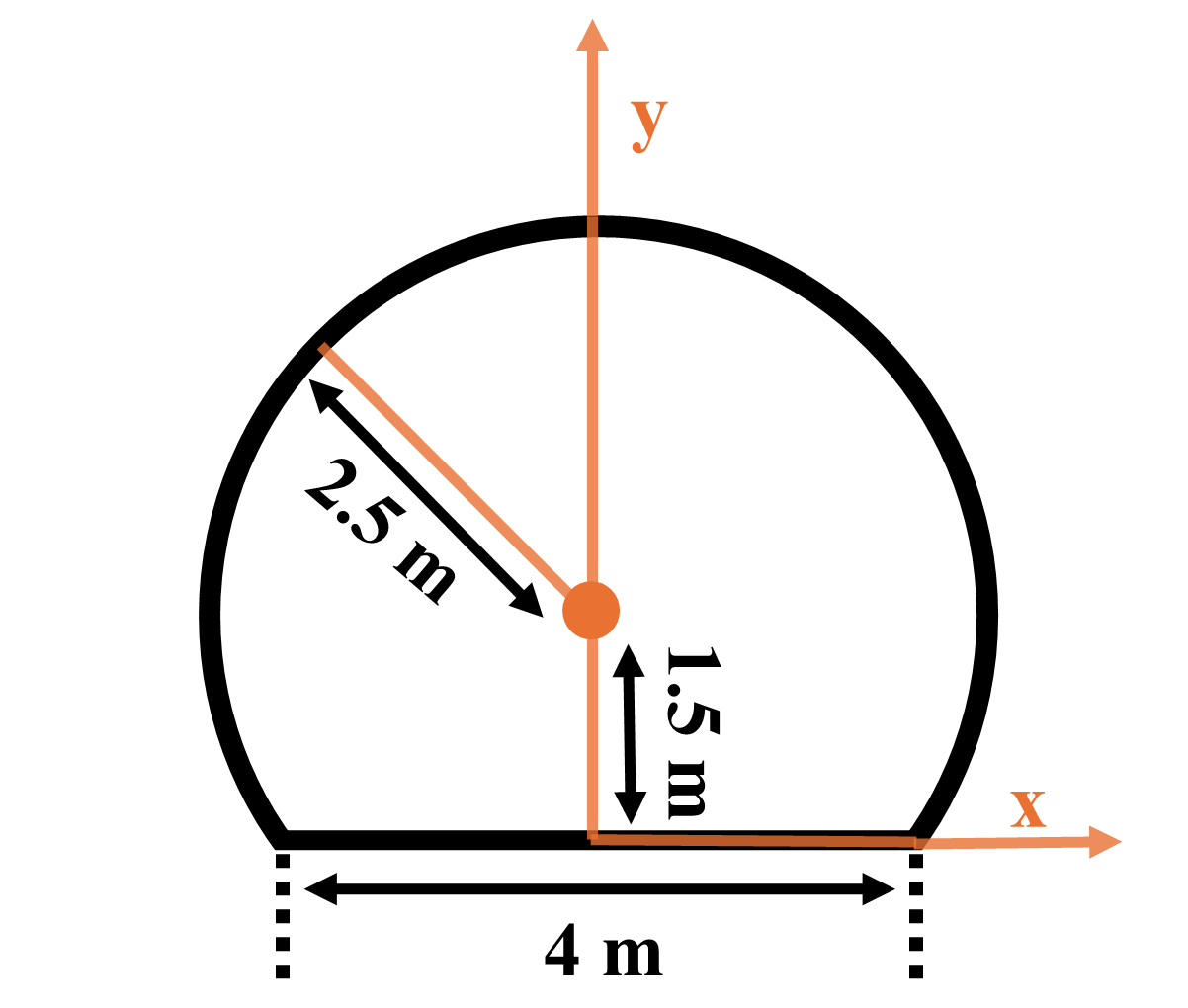}\label{fig:tunnelshape_b}}
\hfill
\subfloat[]{\includegraphics[width=0.24\textwidth]{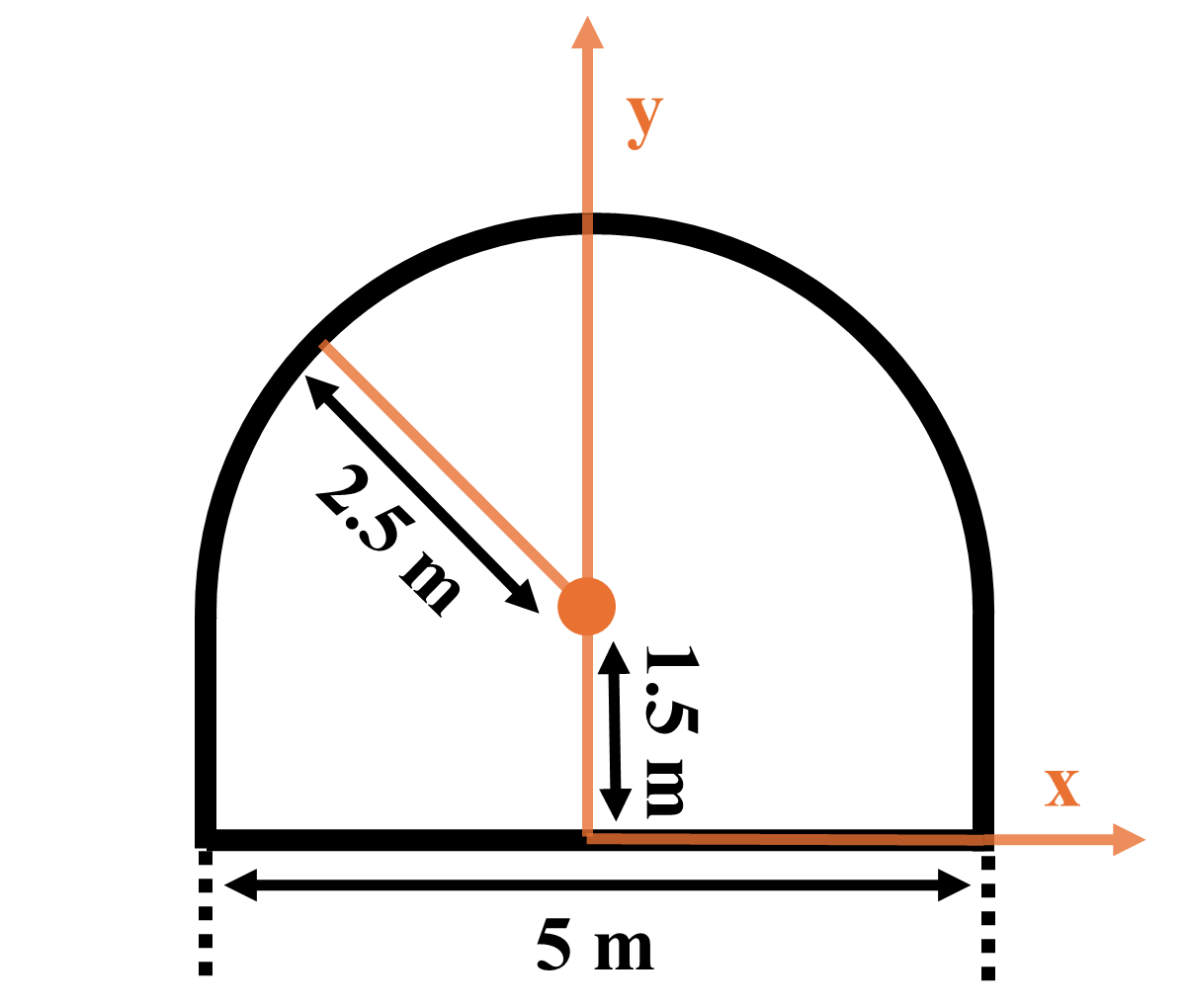}\label{fig:tunnelshape_c}}
\hfill
\subfloat[]{\includegraphics[width=0.24\textwidth]{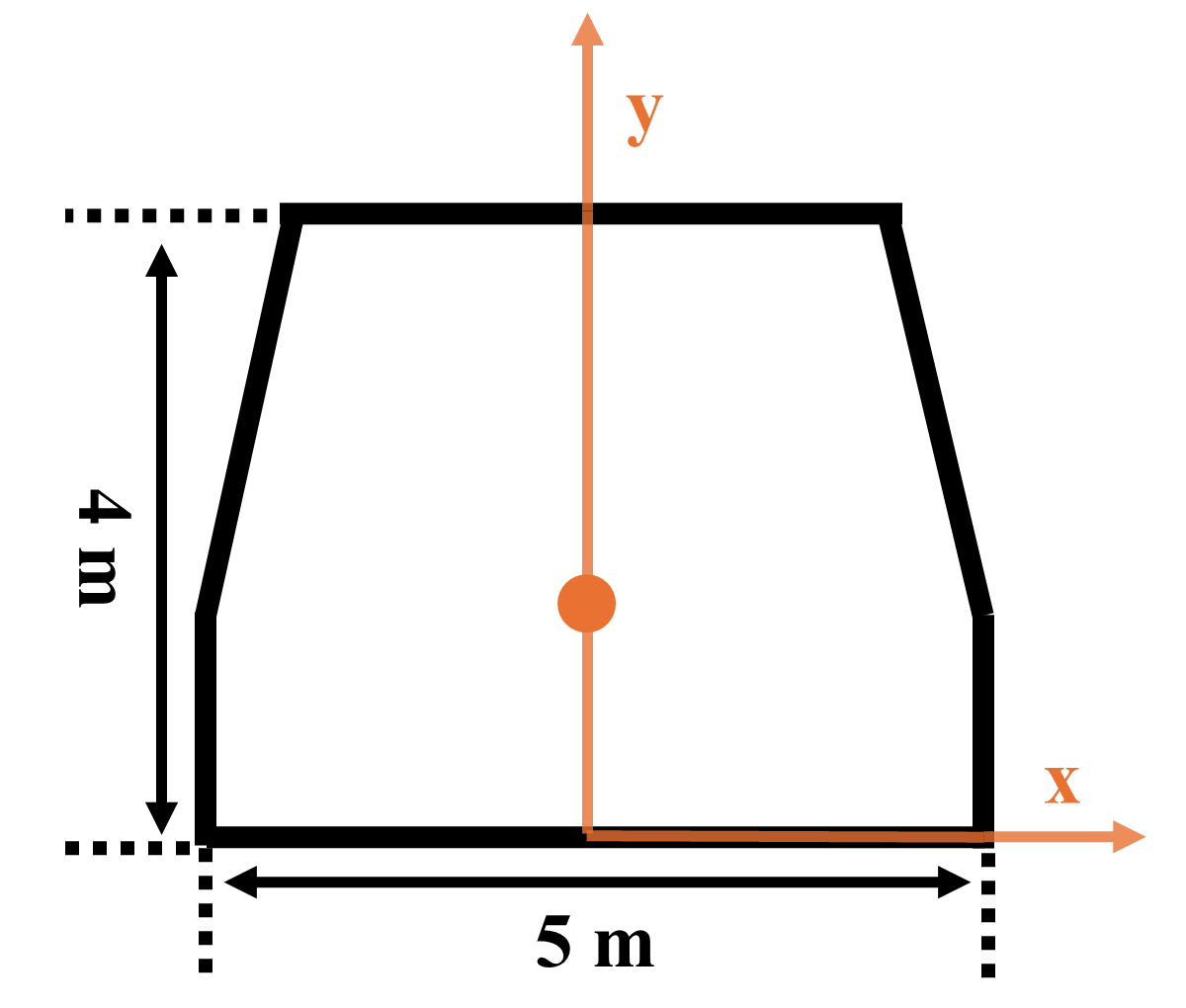}\label{fig:tunnelshape_d}}
\caption{Cross-sectional geometries of the four tunnel types: (a)rectangular, (b) arched, (c) arched but with vertical side walls, and (d) trapezoidal.}
\label{tunnelshape}
\end{figure*}

\begin{figure*}[ht]
	\centering
	\subfloat[Coarse-mesh PWE]{\includegraphics[width=0.24\textwidth]{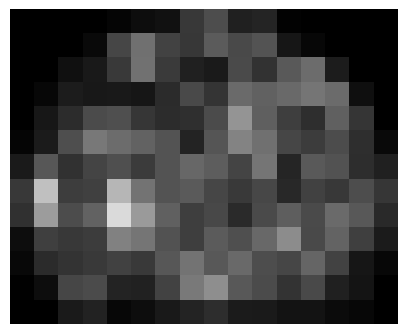}\label{fig:abl_a}} \hfill
	\subfloat[PRBPN without DWTF]{\includegraphics[width=0.24\textwidth]{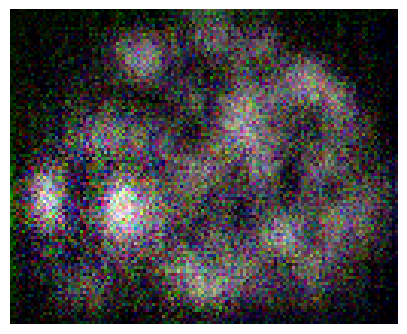}\label{fig:abl_b}} \hfill
	\subfloat[PRBPN]{\includegraphics[width=0.24\textwidth]{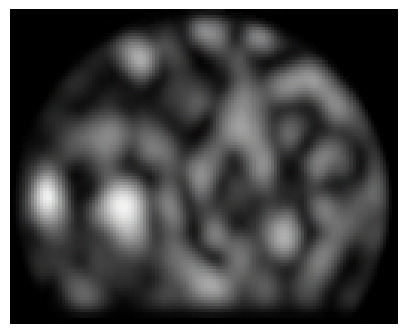}\label{fig:abl_c}} \hfill
	\subfloat[Fine-mesh PWE]{\includegraphics[width=0.24\textwidth]{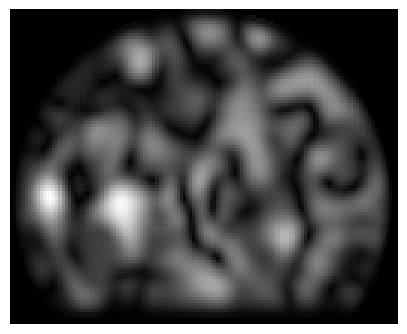}\label{fig:abl_d}} \\
	
	\subfloat[Coarse-mesh PWE]{\includegraphics[width=0.24\textwidth]{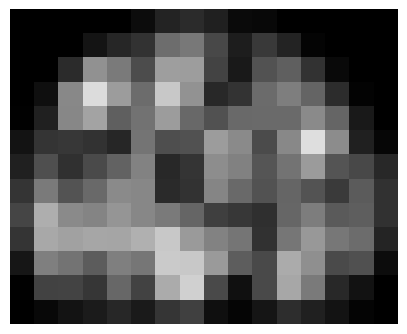}\label{fig:abl_e}} \hfill
	\subfloat[PRBPN without DWTF]{\includegraphics[width=0.24\textwidth]{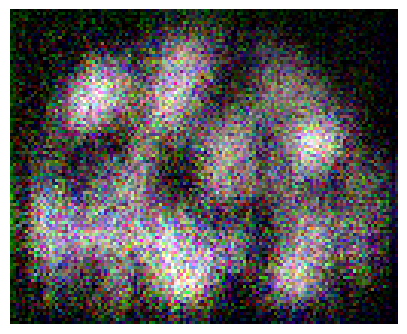}\label{fig:abl_f}} \hfill
	\subfloat[PRBPN]{\includegraphics[width=0.24\textwidth]{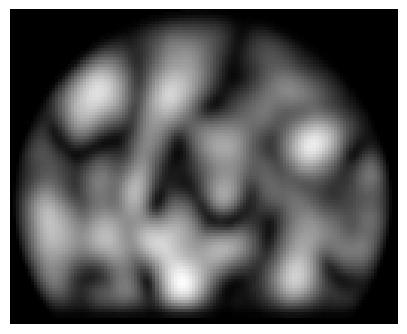}\label{fig:abl_g}} \hfill
	\subfloat[Fine-mesh PWE]{\includegraphics[width=0.24\textwidth]{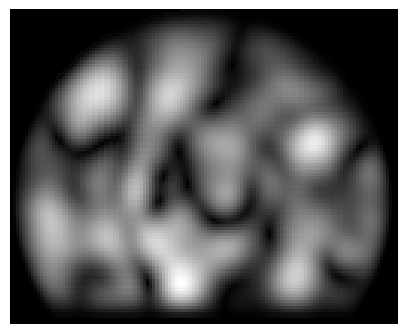}\label{fig:abl_h}}
	
	\caption{Ablation study on DWTF in an arched tunnel: (a) Coarse-mesh PWE at 500m, (b) PRBPN without DWTF at 500m, (c) PRBPN at 500m (d) Fine-mesh PWE at 500m, (e) Coarse-mesh PWE at 1000m, (f) PRBPN without DWTF at 1000m, (g) PRBPN at 1000m and (h) Fine-mesh PWE at 1000m distances.}
	\label{fig:ablation}
\end{figure*}

\begin{table}[H]
\centering
\caption{Setup of the input parameters for the tunnel propagation simulations.}
\label{TableI}
\setlength{\tabcolsep}{3pt}
\renewcommand{\arraystretch}{1.1}
\begin{tabular}{|c|c|c|c|}
\hline
Parameter & Min.\ value & Max.\ value & Increment \\
\hline
shape                  & 1      & 4      & 1     \\
$f$ [GHz]              & 0.9    & 5.9    & 1.0   \\
$\epsilon_r$           & 5      & 10     & 2.5   \\
$\sigma$ [S/m]         & \multicolumn{3}{c|}{\{0.001, 0.01, 0.1\}} \\
$x_{\text{TX}}$ [m]    & 0      & 2.0    & 0.2   \\
$y_{\text{TX}}$ [m]    & 0.5    & 2.5    & 0.5   \\
$x_{\text{RX}}$ [m]    & $-1.5$ & 1.5    & 0.15  \\
$y_{\text{RX}}$ [m]    & 0.2    & 3.2    & 0.15  \\
\hline
\end{tabular}
\end{table}

\begin{figure*}[ht]
\centering
\subfloat[Coarse-mesh PWE]{\includegraphics[width=0.16\textwidth]{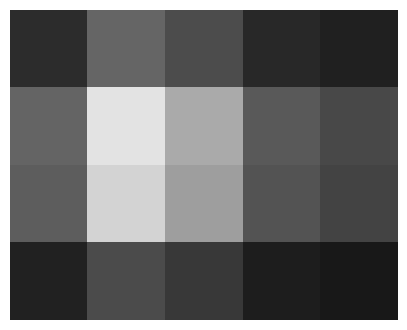}}\label{fig:input500_109} \hfill
\subfloat[PRBPN]{\includegraphics[width=0.16\textwidth]{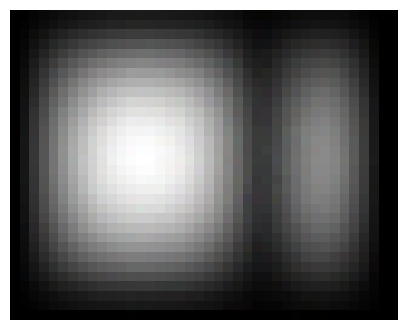}\label{fig:prediction500_109}} \hfill
\subfloat[Fine-mesh PWE]{\includegraphics[width=0.16\textwidth]{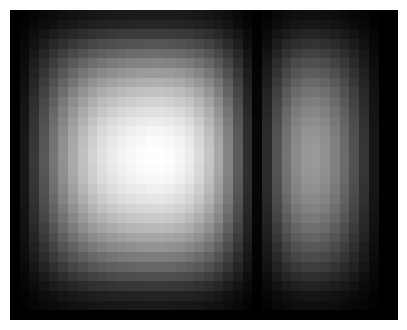}\label{fig:groundtruth500_109}} \hfill
\subfloat[Coarse-mesh PWE]{\includegraphics[width=0.16\textwidth]{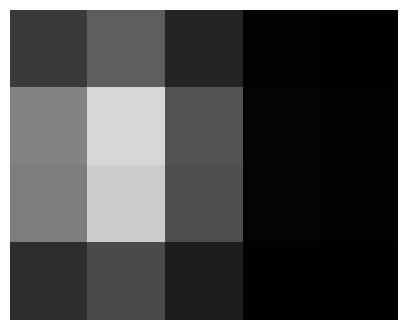}\label{fig:input1000_109}} \hfill
\subfloat[PRBPN]{\includegraphics[width=0.16\textwidth]{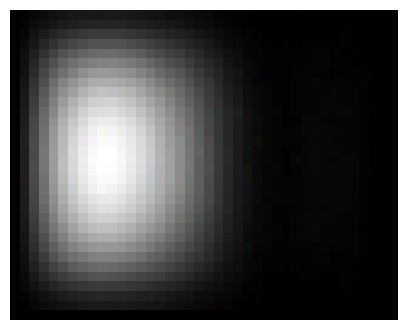}\label{fig:prediction1000_109}} \hfill
\subfloat[Fine-mesh PWE]{\includegraphics[width=0.16\textwidth]{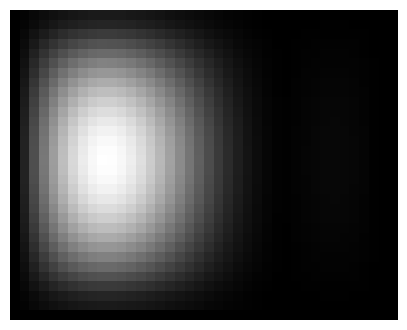}\label{fig:groundtruth1000_109}} \hfill

\caption{Comparison of the results for rectangular tunnel under 0.9G: (a) Coarse-mesh PWE at 500m, (b) PRBPN at 500m, (c) Fine-mesh PWE at 500m, (d) Coarse-mesh PWE at 1000m, (e) PRBPN at 1000m, and (f) Fine-mesh PWE at 1000m distances.}
\label{fig:shape109}
\end{figure*}

\begin{figure*}[ht]
\centering
\subfloat[Coarse-mesh PWE]{\includegraphics[width=0.16\textwidth]{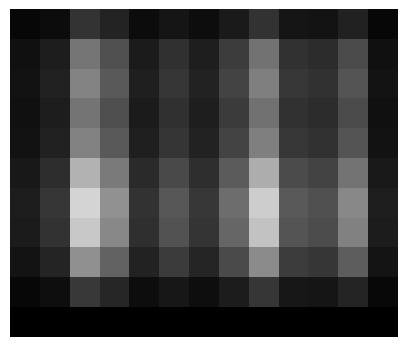}\label{fig:input500_124}} \hfill
\subfloat[PRBPN]{\includegraphics[width=0.16\textwidth]{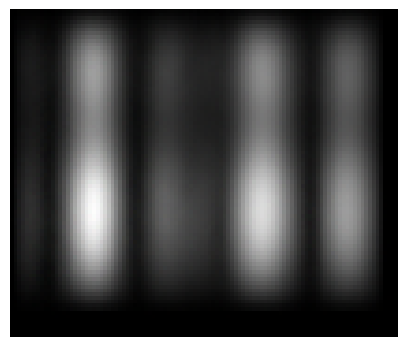}\label{fig:prediction500_124}} \hfill
\subfloat[Fine-mesh PWE]{\includegraphics[width=0.16\textwidth]{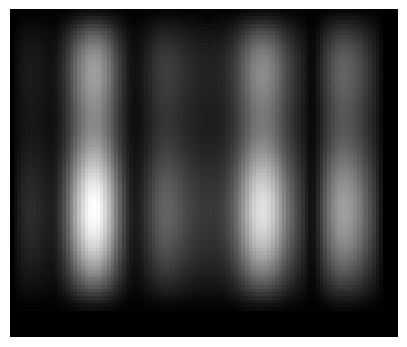}\label{fig:groundtruth500_124}} \hfill
\subfloat[Coarse-mesh PWE]{\includegraphics[width=0.16\textwidth]{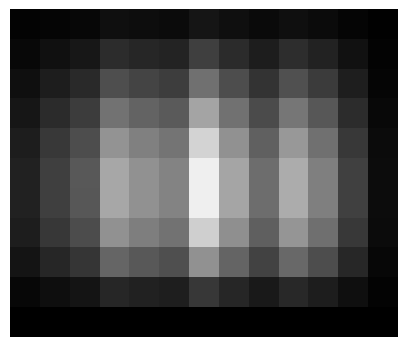}\label{fig:input1000_124}} \hfill
\subfloat[PRBPN]{\includegraphics[width=0.16\textwidth]{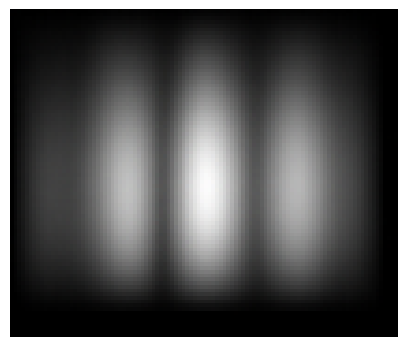}\label{fig:prediction1000_124}} \hfill
\subfloat[Fine-mesh PWE]{\includegraphics[width=0.16\textwidth]{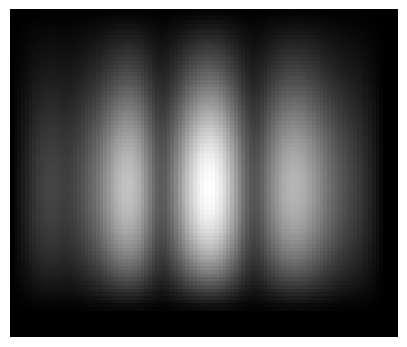}\label{fig:groundtruth1000_124}} \hfill

\caption{Comparison of the results for rectangular tunnel under 2.4G: (a) Coarse-mesh PWE at 500m, (b) PRBPN at 500m, (c) Fine-mesh PWE at 500m, (d) Coarse-mesh PWE at 1000m, (e) PRBPN at 1000m, and (f) Fine-mesh PWE at 1000m distances.}
\label{fig:shape124}
\end{figure*}

\begin{figure*}[ht]
\centering
\subfloat[Coarse-mesh PWE]{\includegraphics[width=0.16\textwidth]{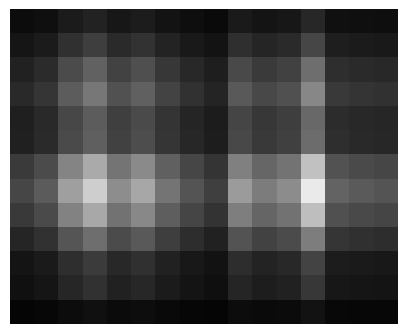}\label{fig:input500_149}} \hfill
\subfloat[PRBPN]{\includegraphics[width=0.16\textwidth]{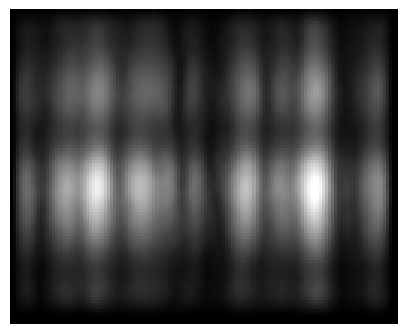}\label{fig:prediction500_149}} \hfill
\subfloat[Fine-mesh PWE]{\includegraphics[width=0.16\textwidth]{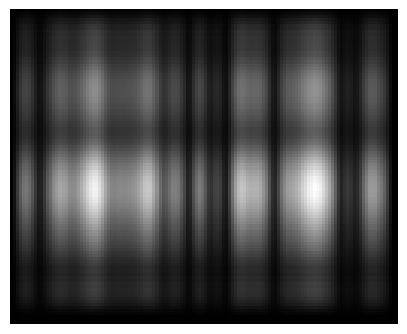}\label{fig:groundtruth500_149}} \hfill
\subfloat[Coarse-mesh PWE]{\includegraphics[width=0.16\textwidth]{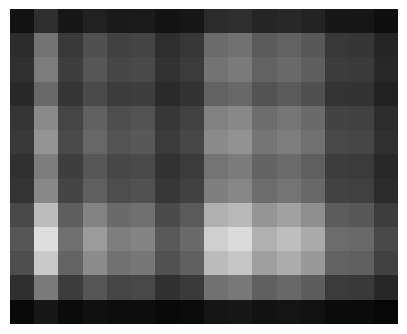}\label{fig:input1000_149}} \hfill
\subfloat[PRBPN]{\includegraphics[width=0.16\textwidth]{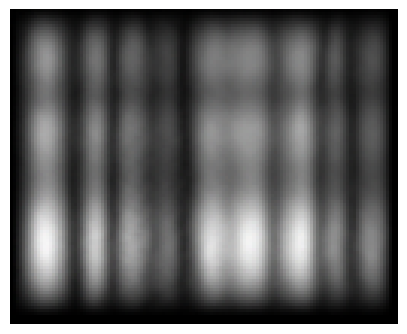}\label{fig:prediction1000_149}} \hfill
\subfloat[Fine-mesh PWE]{\includegraphics[width=0.16\textwidth]{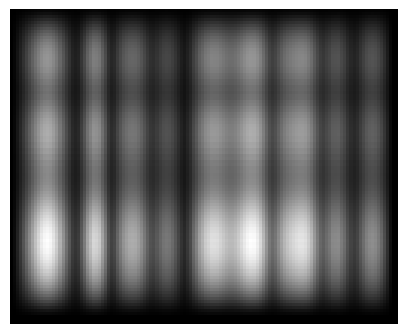}\label{fig:groundtruth1000_149}} \hfill

\caption{Comparison of the results for rectangular tunnel under 4.9G: (a) Coarse-mesh PWE at 500m, (b) PRBPN at 500m, (c) Fine-mesh PWE at 500m, (d) Coarse-mesh PWE at 1000m, (e) PRBPN at 1000m, and (f) Fine-mesh PWE at 1000m distances.}
\label{fig:shape149}
\end{figure*}

\begin{figure*}[ht]
\centering
\subfloat[Coarse-mesh PWE]{\includegraphics[width=0.16\textwidth]{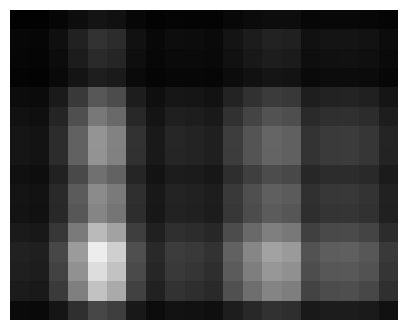}\label{fig:input500_158}} \hfill
\subfloat[PRBPN]{\includegraphics[width=0.16\textwidth]{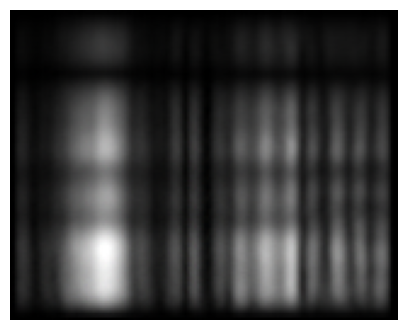}\label{fig:prediction500_158}} \hfill
\subfloat[Fine-mesh PWE]{\includegraphics[width=0.16\textwidth]{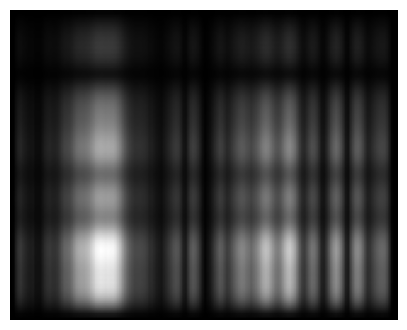}\label{fig:groundtruth500_158}} \hfill
\subfloat[Coarse-mesh PWE]{\includegraphics[width=0.16\textwidth]{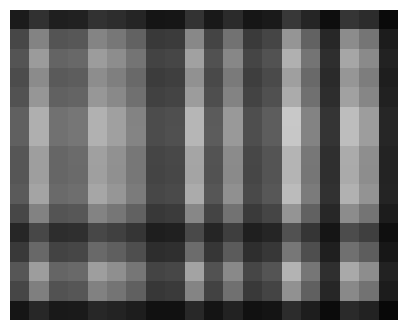}\label{fig:input1000_158}} \hfill
\subfloat[PRBPN]{\includegraphics[width=0.16\textwidth]{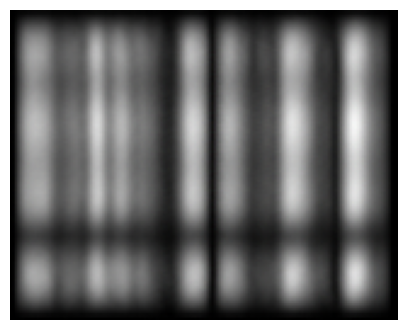}\label{fig:prediction1000_158}} \hfill
\subfloat[Fine-mesh PWE]{\includegraphics[width=0.16\textwidth]{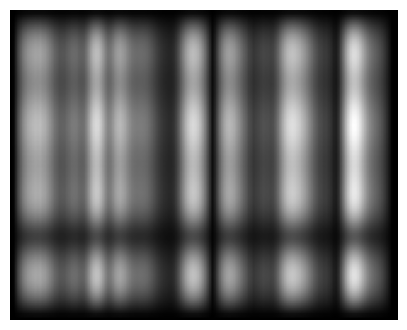}\label{fig:groundtruth1000_158}} \hfill

\caption{Comparison of the results for rectangular tunnel under 5.8G: (a) Coarse-mesh PWE at 500m, (b) PRBPN at 500m, (c) Fine-mesh PWE at 500m, (d) Coarse-mesh PWE at 1000m, (e) PRBPN at 1000m, and (f) Fine-mesh PWE at 1000m distances.}
\label{fig:shape158}
\end{figure*}


\section{Numerical Examples}\label{Section III}
\subsection{Simulation and Dataset Generation}

To validate the proposed PRBPN model, we conduct numerical simulations using the conventional PWE method over a \(1000\,\mathrm{m}\) tunnel under varying boundary and antenna configurations. Four representative tunnel geometries with dielectric, lossy walls are considered for training: a rectangular tunnel (shape~1), an arched tunnel (shape~2), an arched tunnel with vertical side walls (shape~3), and a trapezoidal tunnel (shape~4). The corresponding cross sections are shown in Fig.~\ref{tunnelshape}(a)–(d).
At the initial cross section, a Gaussian beam with unit amplitude is launched from a transmitting antenna at \((x_{\mathrm{TX}}, y_{\mathrm{TX}})\), with a beam width set to a standard deviation of \(3\lambda\).

For the PWE simulations, the coarse-mesh configuration uses \(\Delta x=\Delta y=3.2\lambda\) and \(\Delta z=2\lambda\), while the fine-mesh configuration uses \(\Delta x=\Delta y=0.4\lambda\) and \(\Delta z=2\lambda\), yielding a volumetric resolution ratio of 1:64. To minimize bias from transmitter placement and to rigorously assess generalization, we partition antenna locations into disjoint training and testing regimes: training uses tunnels with \(\mathrm{TX1}\in[0,1.0]\) and \(\mathrm{TX2}\in[0,1.0]\), whereas testing uses \(\mathrm{TX1}\in[1.2,2.0]\) and \(\mathrm{TX2}\in[2.0,2.5]\). 
This separation prevents data leakage across spatial neighborhoods and allows evaluation on more challenging edge-proximal configurations. In these cases, larger offsets position the emitter closer to the side walls, increasing grazing incidence, wall coupling, and higher-order mode excitation. Such conditions accentuate multipath effects and attenuation asymmetries.
These settings provide an efficient yet physically consistent resolution of wavefront evolution along the tunnel. The ranges and increments of all input parameters, including wall material properties, antenna positions, and tunnel shapes, are summarized in Table~\ref{TableI}. The dataset is assembled by exhaustive sampling over these combinations to ensure broad coverage and diversity of tunnel propagation scenarios.

\begin{figure*}[ht]
\centering
\subfloat[Coarse-mesh PWE]{\includegraphics[width=0.16\textwidth]{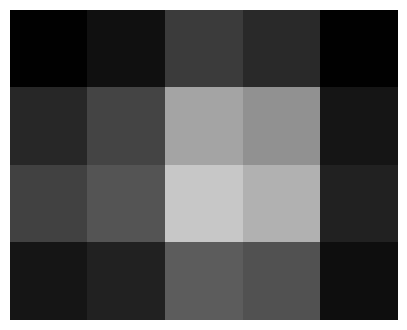}\label{fig:input500_209}} \hfill
\subfloat[PRBPN]{\includegraphics[width=0.16\textwidth]{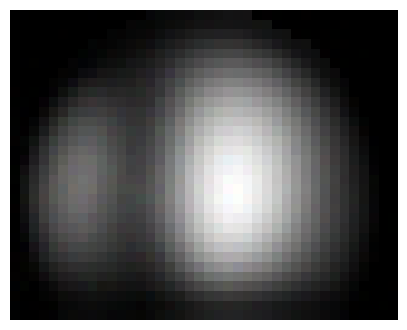}\label{fig:prediction500_209}} \hfill
\subfloat[Fine-mesh PWE]{\includegraphics[width=0.16\textwidth]{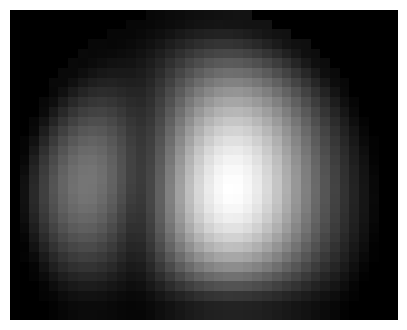}\label{fig:groundtruth500_209}} \hfill
\subfloat[Coarse-mesh PWE]{\includegraphics[width=0.16\textwidth]{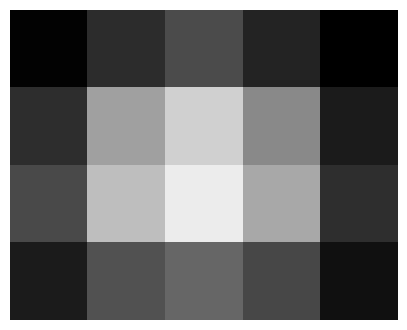}\label{fig:input1000_209}} \hfill
\subfloat[PRBPN]{\includegraphics[width=0.16\textwidth]{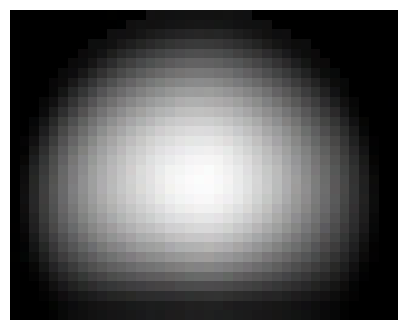}\label{fig:prediction1000_209}} \hfill
\subfloat[Fine-mesh PWE]{\includegraphics[width=0.16\textwidth]{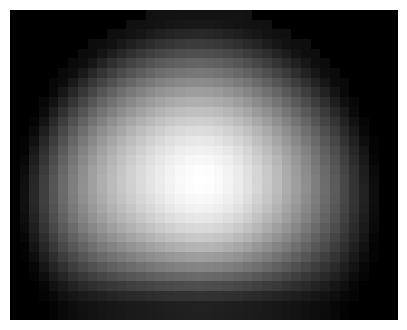}\label{fig:groundtruth1000_209}} \hfill

\caption{Comparison of the results for arched tunnel under 0.9G: (a) Coarse-mesh PWE at 500m, (b) PRBPN at 500m, (c) Fine-mesh PWE at 500m, (d) Coarse-mesh PWE at 1000m, (e) PRBPN at 1000m, and (f) Fine-mesh PWE at 1000m distances.}
\label{fig:shape209}
\end{figure*}

\begin{figure*}[ht]
\centering
\subfloat[Coarse-mesh PWE]{\includegraphics[width=0.16\textwidth]{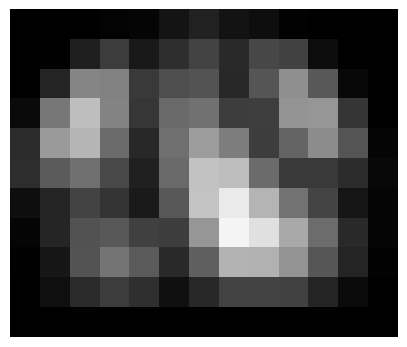}\label{fig:input500_224}} \hfill
\subfloat[PRBPN]{\includegraphics[width=0.16\textwidth]{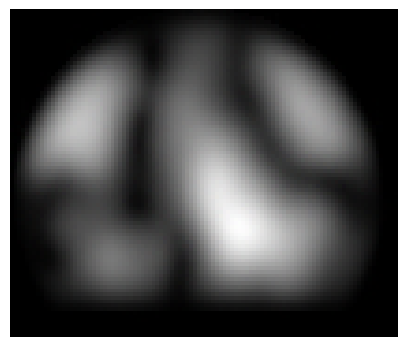}\label{fig:prediction500_224}} \hfill
\subfloat[Fine-mesh PWE]{\includegraphics[width=0.16\textwidth]{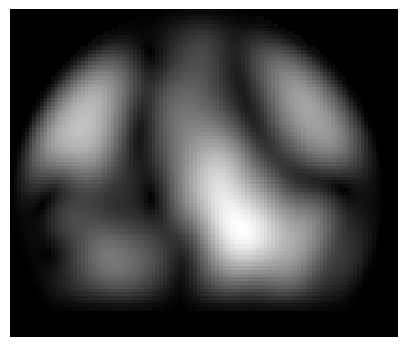}\label{fig:groundtruth500_224}} \hfill
\subfloat[Coarse-mesh PWE]{\includegraphics[width=0.16\textwidth]{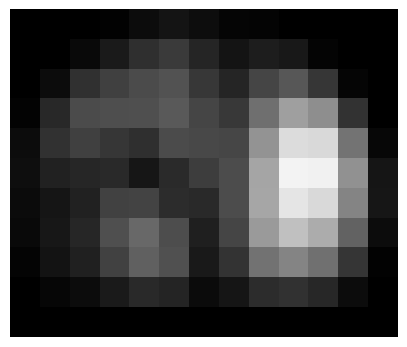}\label{fig:input1000_224}} \hfill
\subfloat[PRBPN]{\includegraphics[width=0.16\textwidth]{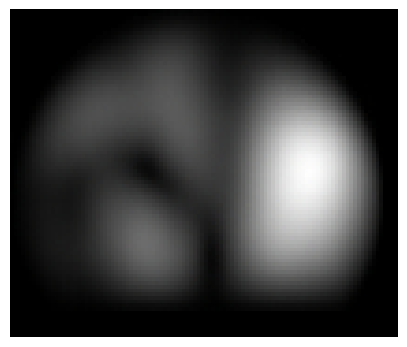}\label{fig:prediction1000_224}} \hfill
\subfloat[Fine-mesh PWE]{\includegraphics[width=0.16\textwidth]{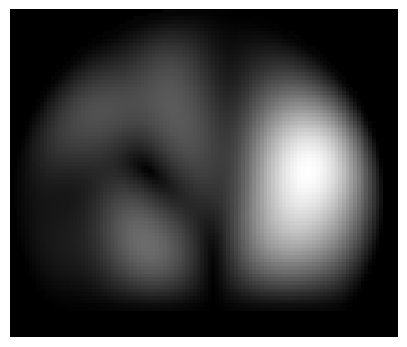}\label{fig:groundtruth1000_224}} \hfill

\caption{Comparison of the results for arched tunnel under 2.4G: (a) Coarse-mesh PWE at 500m, (b) PRBPN at 500m, (c) Fine-mesh PWE at 500m, (d) Coarse-mesh PWE at 1000m, (e) PRBPN at 1000m, and (f) Fine-mesh PWE at 1000m distances.}
\label{fig:shape224}
\end{figure*}

\begin{figure*}[ht]
\centering
\subfloat[Coarse-mesh PWE]{\includegraphics[width=0.16\textwidth]{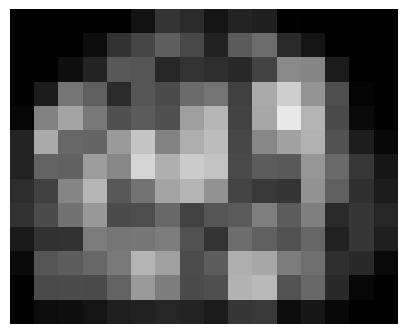}\label{fig:input500_249}} \hfill
\subfloat[PRBPN]{\includegraphics[width=0.16\textwidth]{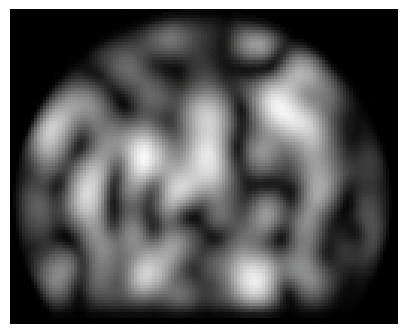}\label{fig:prediction500_249}} \hfill
\subfloat[Fine-mesh PWE]{\includegraphics[width=0.16\textwidth]{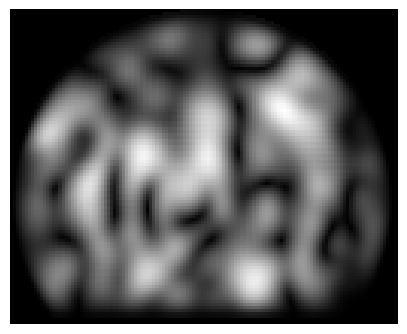}\label{fig:groundtruth500_249}} \hfill
\subfloat[Coarse-mesh PWE]{\includegraphics[width=0.16\textwidth]{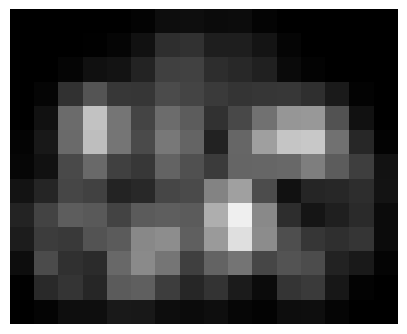}\label{fig:input1000_249}} \hfill
\subfloat[PRBPN]{\includegraphics[width=0.16\textwidth]{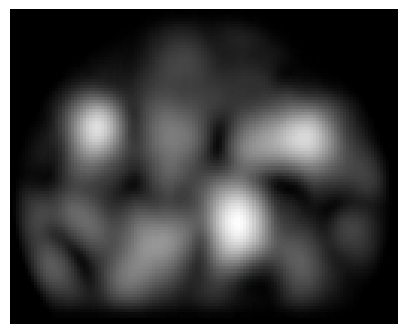}\label{fig:prediction1000_249}} \hfill
\subfloat[Fine-mesh PWE]{\includegraphics[width=0.16\textwidth]{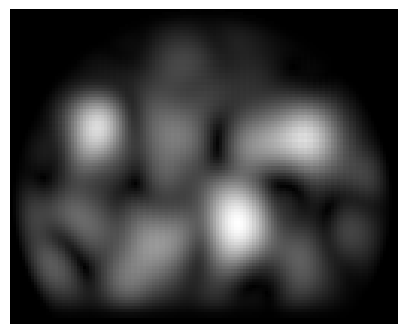}\label{fig:groundtruth1000_249}} \hfill

\caption{Comparison of the results for arched tunnel under 4.9G: (a) Coarse-mesh PWE at 500m, (b) PRBPN at 500m, (c) Fine-mesh PWE at 500m, (d) Coarse-mesh PWE at 1000m, (e) PRBPN at 1000m, and (f) Fine-mesh PWE at 1000m distances.}
\label{fig:shape249}
\end{figure*}

\begin{figure*}[ht]
\centering
\subfloat[Coarse-mesh PWE]{\includegraphics[width=0.16\textwidth]{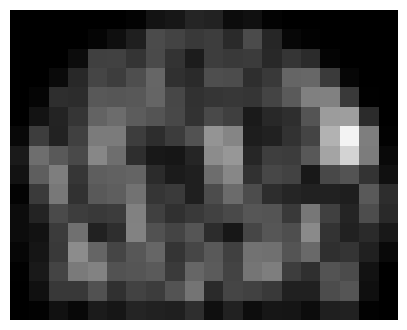}\label{fig:input500_258}} \hfill
\subfloat[PRBPN]{\includegraphics[width=0.16\textwidth]{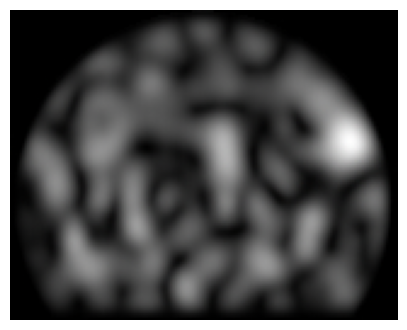}\label{fig:prediction500_258}} \hfill
\subfloat[Fine-mesh PWE]{\includegraphics[width=0.16\textwidth]{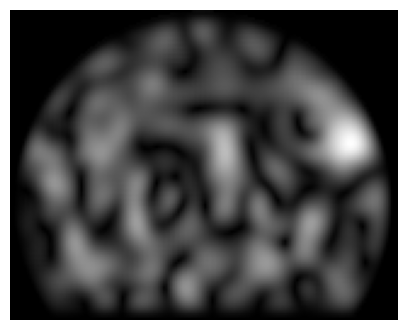}\label{fig:groundtruth500_258}} \hfill
\subfloat[Coarse-mesh PWE]{\includegraphics[width=0.16\textwidth]{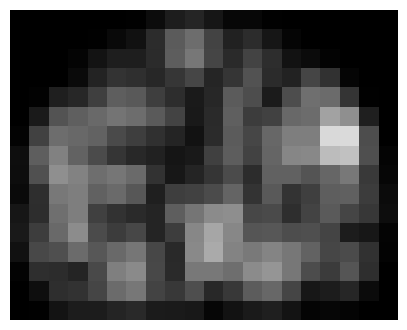}\label{fig:input1000_258}} \hfill
\subfloat[PRBPN]{\includegraphics[width=0.16\textwidth]{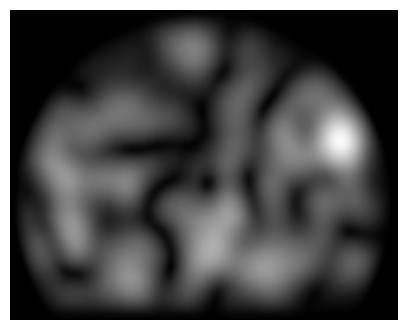}\label{fig:prediction1000_258}} \hfill
\subfloat[Fine-mesh PWE]{\includegraphics[width=0.16\textwidth]{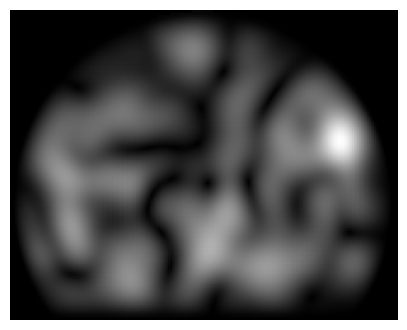}\label{fig:groundtruth1000_258}} \hfill

\caption{Comparison of the results for arched tunnel under 5.8G: (a) Coarse-mesh PWE at 500m, (b) PRBPN at 500m, (c) Fine-mesh PWE at 500m, (d) Coarse-mesh PWE at 1000m, (e) PRBPN at 1000m, and (f) Fine-mesh PWE at 1000m distances.}
\label{fig:shape258}
\end{figure*}

\begin{figure*}[ht]
\centering
\subfloat[Coarse-mesh PWE]{\includegraphics[width=0.16\textwidth]{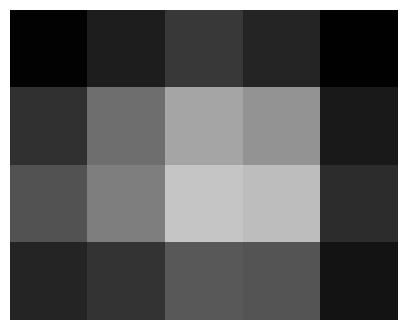}\label{fig:input500_309}} \hfill
\subfloat[PRBPN]{\includegraphics[width=0.16\textwidth]{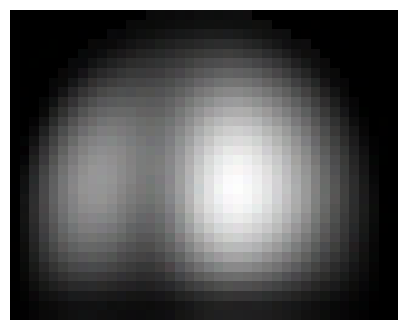}\label{fig:prediction500_309}} \hfill
\subfloat[Fine-mesh PWE]{\includegraphics[width=0.16\textwidth]{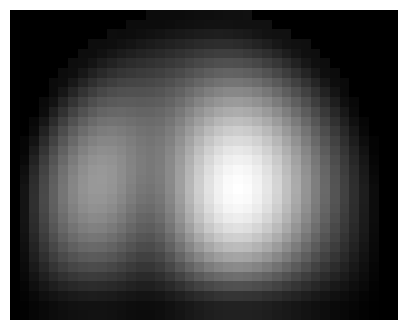}\label{fig:groundtruth500_309}} \hfill
\subfloat[Coarse-mesh PWE]{\includegraphics[width=0.16\textwidth]{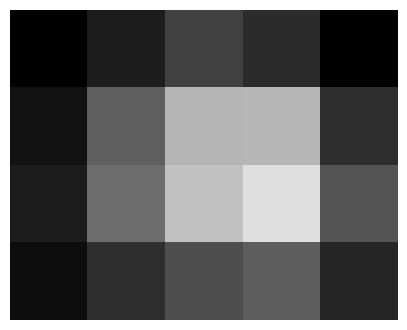}\label{fig:input1000_309}} \hfill
\subfloat[PRBPN]{\includegraphics[width=0.16\textwidth]{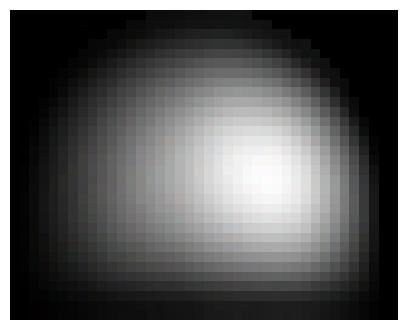}\label{fig:prediction1000_309}} \hfill
\subfloat[Fine-mesh PWE]{\includegraphics[width=0.16\textwidth]{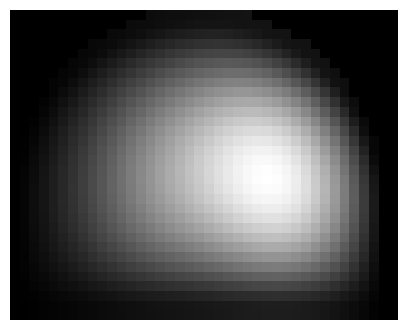}\label{fig:groundtruth1000_309}} \hfill

\caption{Comparison of the results for arched but with vertical side walls tunnel with vertical side walls under 4.9G: (a) Coarse-mesh PWE at 500m, (b) PRBPN at 500m, (c) Fine-mesh PWE at 500m, (d) Coarse-mesh PWE at 1000m, (e) PRBPN at 1000m, and (f) Fine-mesh PWE at 1000m distances.}
\label{fig:shape309}
\end{figure*}

\begin{figure*}[ht]
\centering
\subfloat[Coarse-mesh PWE]{\includegraphics[width=0.16\textwidth]{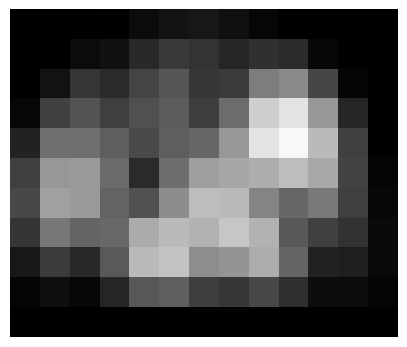}\label{fig:input500_324}} \hfill
\subfloat[PRBPN]{\includegraphics[width=0.16\textwidth]{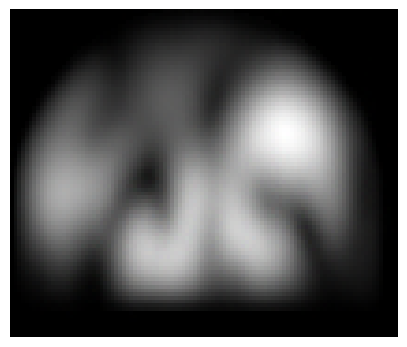}\label{fig:prediction500_324}} \hfill
\subfloat[Fine-mesh PWE]{\includegraphics[width=0.16\textwidth]{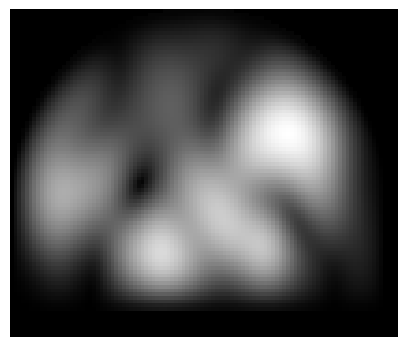}\label{fig:groundtruth500_324}} \hfill
\subfloat[Coarse-mesh PWE]{\includegraphics[width=0.16\textwidth]{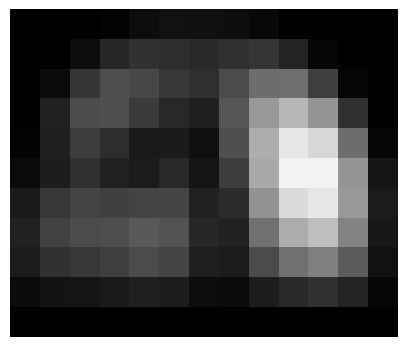}\label{fig:input1000_324}} \hfill
\subfloat[PRBPN]{\includegraphics[width=0.16\textwidth]{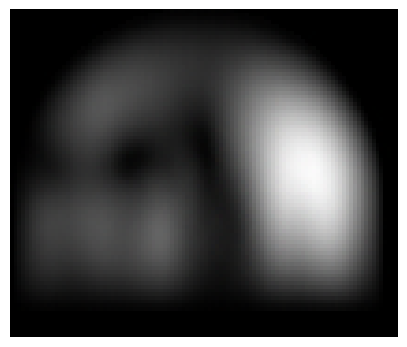}\label{fig:prediction1000_324}} \hfill
\subfloat[Fine-mesh PWE]{\includegraphics[width=0.16\textwidth]{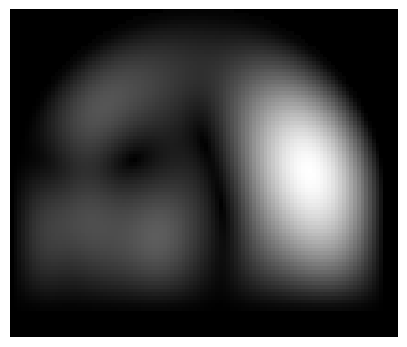}\label{fig:groundtruth1000_324}} \hfill

\caption{Comparison of the results for arched tunnel with vertical side walls under 2.4G: (a) Coarse-mesh PWE at 500m, (b) PRBPN at 500m, (c) Fine-mesh PWE at 500m, (d) Coarse-mesh PWE at 1000m, (e) PRBPN at 1000m, and (f) Fine-mesh PWE at 1000m distances.}
\label{fig:shape324}
\end{figure*}

\begin{figure*}[ht]
\centering
\subfloat[Coarse-mesh PWE]{\includegraphics[width=0.16\textwidth]{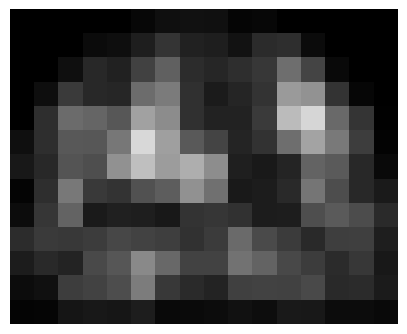}\label{fig:input500_349}} \hfill
\subfloat[PRBPN]{\includegraphics[width=0.16\textwidth]{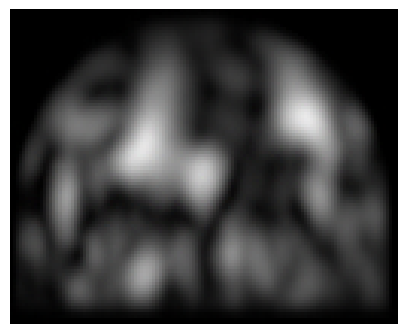}\label{fig:prediction500_349}} \hfill
\subfloat[Fine-mesh PWE]{\includegraphics[width=0.16\textwidth]{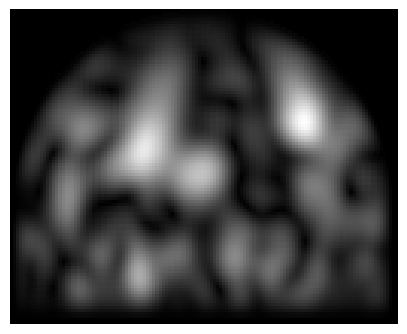}\label{fig:groundtruth500_349}} \hfill
\subfloat[Coarse-mesh PWE]{\includegraphics[width=0.16\textwidth]{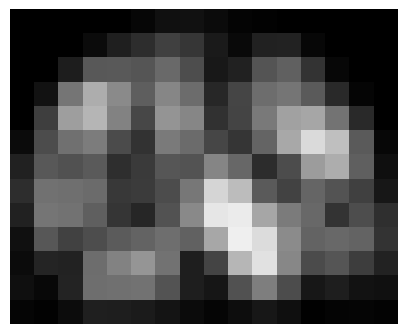}\label{fig:input1000_349}} \hfill
\subfloat[PRBPN]{\includegraphics[width=0.16\textwidth]{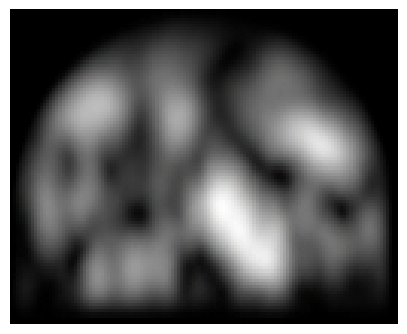}\label{fig:prediction1000_349}} \hfill
\subfloat[Fine-mesh PWE]{\includegraphics[width=0.16\textwidth]{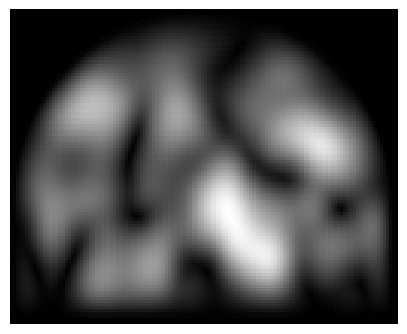}\label{fig:groundtruth1000_349}} \hfill

\caption{Comparison of the results for arched tunnel with vertical side walls under 4.9G: (a) Coarse-mesh PWE at 500m, (b) PRBPN at 500m, (c) Fine-mesh PWE at 500m, (d) Coarse-mesh PWE at 1000m, (e) PRBPN at 1000m, and (f) Fine-mesh PWE at 1000m distances.}
\label{fig:shape349}
\end{figure*}

\begin{figure*}[ht]
\centering
\subfloat[Coarse-mesh PWE]{\includegraphics[width=0.16\textwidth]{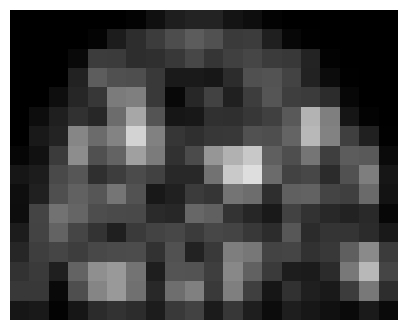}\label{fig:input500_358}} \hfill
\subfloat[PRBPN]{\includegraphics[width=0.16\textwidth]{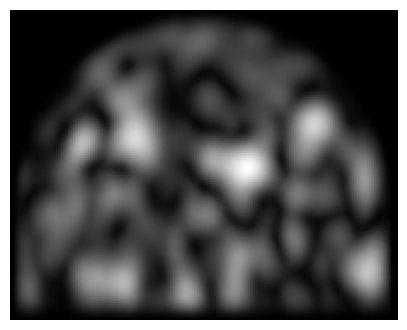}\label{fig:prediction500_358}} \hfill
\subfloat[Fine-mesh PWE]{\includegraphics[width=0.16\textwidth]{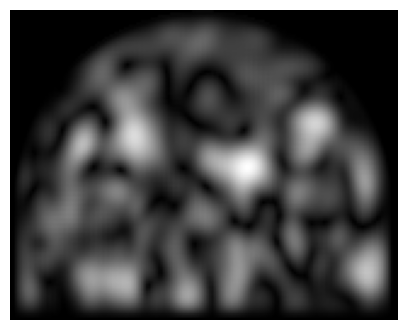}\label{fig:groundtruth500_358}} \hfill
\subfloat[Coarse-mesh PWE]{\includegraphics[width=0.16\textwidth]{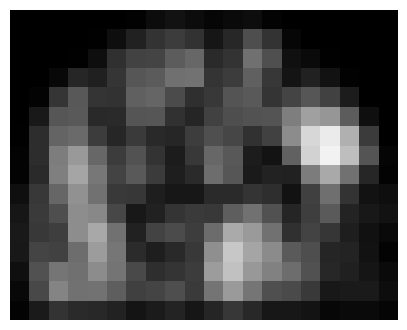}\label{fig:input1000_358}} \hfill
\subfloat[PRBPN]{\includegraphics[width=0.16\textwidth]{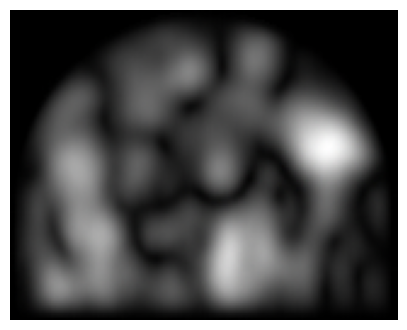}\label{fig:prediction1000_358}} \hfill
\subfloat[Fine-mesh PWE]{\includegraphics[width=0.16\textwidth]{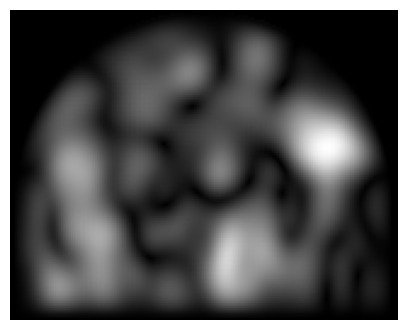}\label{fig:groundtruth1000_358}} \hfill

\caption{Comparison of the results for arched tunnel with vertical side walls under 5.8G: (a) Coarse-mesh PWE at 500m, (b) PRBPN at 500m, (c) Fine-mesh PWE at 500m, (d) Coarse-mesh PWE at 1000m, (e) PRBPN at 1000m, and (f) Fine-mesh PWE at 1000m distances.}
\label{fig:shape358}
\end{figure*}

\begin{figure*}[ht]
\centering
\subfloat[Coarse-mesh PWE]{\includegraphics[width=0.16\textwidth]{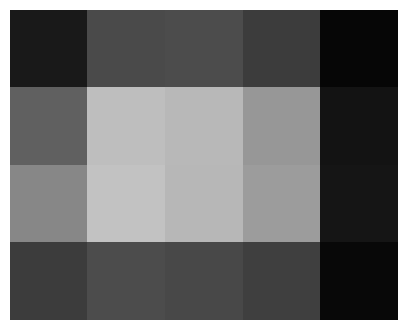}\label{fig:input500_409}} \hfill
\subfloat[PRBPN]{\includegraphics[width=0.16\textwidth]{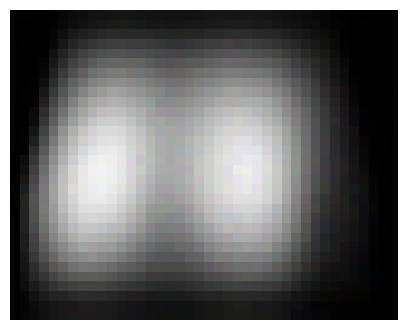}\label{fig:prediction500_409}} \hfill
\subfloat[Fine-mesh PWE]{\includegraphics[width=0.16\textwidth]{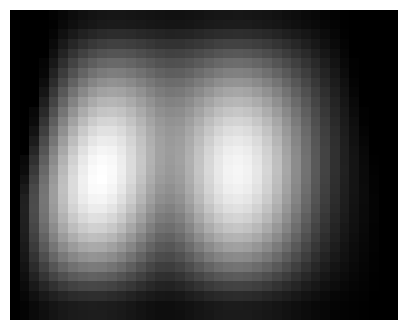}\label{fig:groundtruth500_409}} \hfill
\subfloat[Coarse-mesh PWE]{\includegraphics[width=0.16\textwidth]{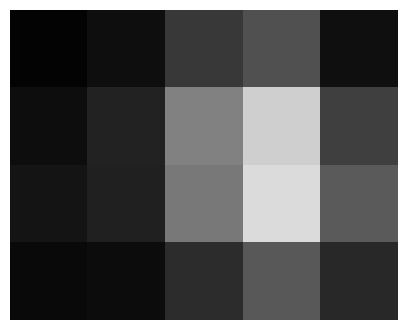}\label{fig:input1000_409}} \hfill
\subfloat[PRBPN]{\includegraphics[width=0.16\textwidth]{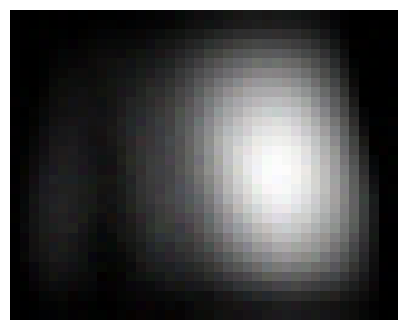}\label{fig:prediction1000_409}} \hfill
\subfloat[Fine-mesh PWE]{\includegraphics[width=0.16\textwidth]{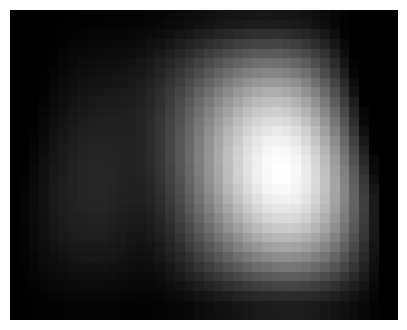}\label{fig:groundtruth1000_409}} \hfill

\caption{Comparison of the results for trapezoidal tunnel under 0.9G: (a) Coarse-mesh PWE at 500m, (b) PRBPN at 500m, (c) Fine-mesh PWE at 500m, (d) Coarse-mesh PWE at 1000m, (e) PRBPN at 1000m, and (f) Fine-mesh PWE at 1000m distances.}
\label{fig:shape409}
\end{figure*}

\begin{figure*}[ht]
\centering
\subfloat[Coarse-mesh PWE]{\includegraphics[width=0.16\textwidth]{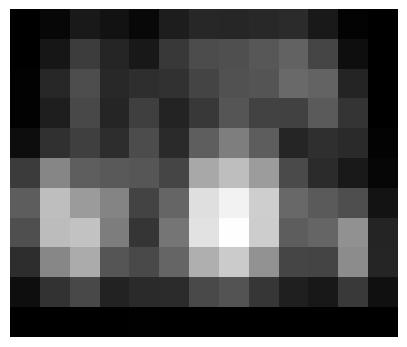}\label{fig:input500_424}} \hfill
\subfloat[PRBPN]{\includegraphics[width=0.16\textwidth]{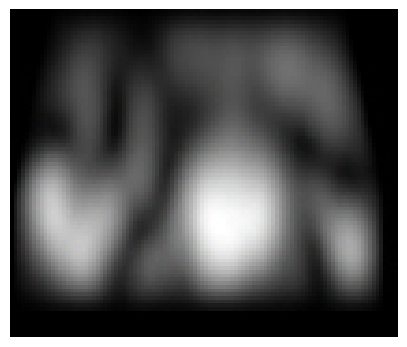}\label{fig:prediction500_424}} \hfill
\subfloat[Fine-mesh PWE]{\includegraphics[width=0.16\textwidth]{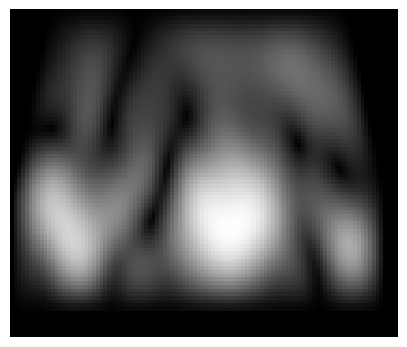}\label{fig:groundtruth500_424}} \hfill
\subfloat[Coarse-mesh PWE]{\includegraphics[width=0.16\textwidth]{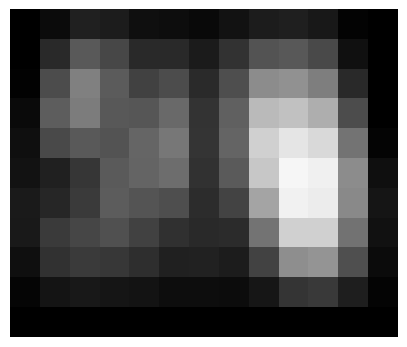}\label{fig:input1000_424}} \hfill
\subfloat[PRBPN]{\includegraphics[width=0.16\textwidth]{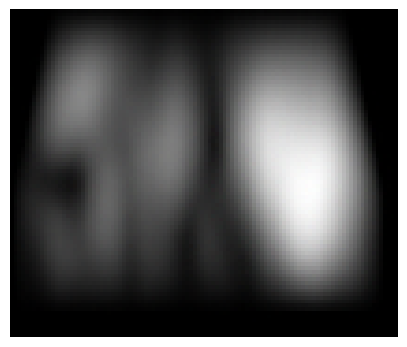}\label{fig:prediction1000_424}} \hfill
\subfloat[Fine-mesh PWE]{\includegraphics[width=0.16\textwidth]{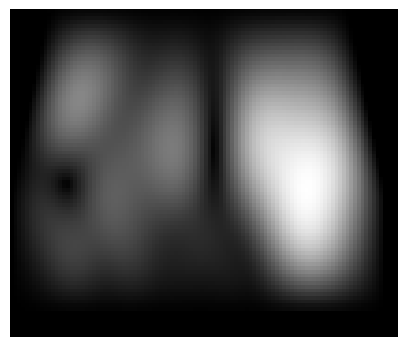}\label{fig:groundtruth1000_424}} \hfill

\caption{Comparison of the results for trapezoidal tunnel under 2.4G: (a) Coarse-mesh PWE at 500m, (b) PRBPN at 500m, (c) Fine-mesh PWE at 500m, (d) Coarse-mesh PWE at 1000m, (e) PRBPN at 1000m, and (f) Fine-mesh PWE at 1000m distances.}
\label{fig:shape424}
\end{figure*}

\begin{figure*}[ht]
\centering
\subfloat[Coarse-mesh PWE]{\includegraphics[width=0.16\textwidth]{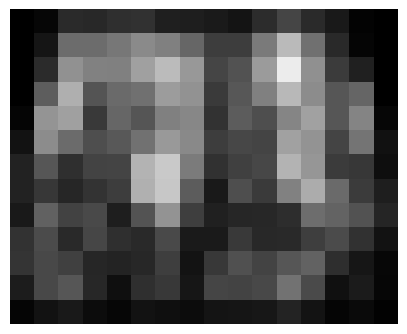}\label{fig:input500_449}} \hfill
\subfloat[PRBPN]{\includegraphics[width=0.16\textwidth]{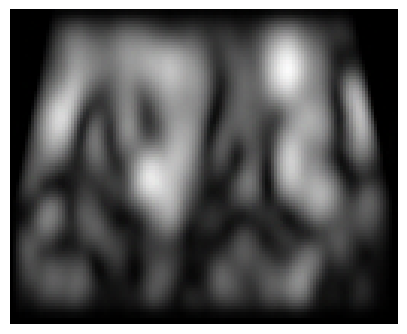}\label{fig:prediction500_449}} \hfill
\subfloat[Fine-mesh PWE]{\includegraphics[width=0.16\textwidth]{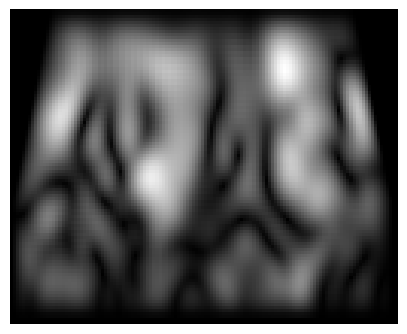}\label{fig:groundtruth500_449}} \hfill
\subfloat[Coarse-mesh PWE]{\includegraphics[width=0.16\textwidth]{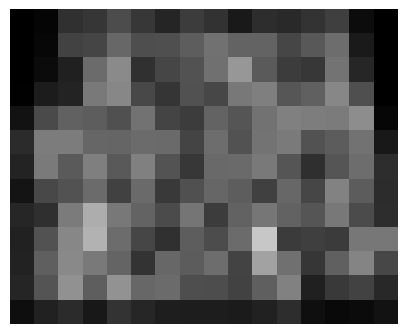}\label{fig:input1000_449}} \hfill
\subfloat[PRBPN]{\includegraphics[width=0.16\textwidth]{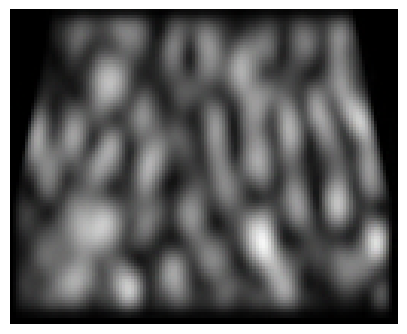}\label{fig:prediction1000_449}} \hfill
\subfloat[Fine-mesh PWE]{\includegraphics[width=0.16\textwidth]{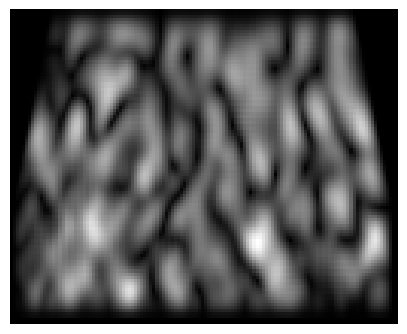}\label{fig:groundtruth1000_449}} \hfill

\caption{Comparison of the results for trapezoidal tunnel under 4.9G: (a) Coarse-mesh PWE at 500m, (b) PRBPN at 500m, (c) Fine-mesh PWE at 500m, (d) Coarse-mesh PWE at 1000m, (e) PRBPN at 1000m, and (f) Fine-mesh PWE at 1000m distances.}
\label{fig:shape449}
\end{figure*}

\begin{figure*}[ht]
\centering
\subfloat[Coarse-mesh PWE]{\includegraphics[width=0.16\textwidth]{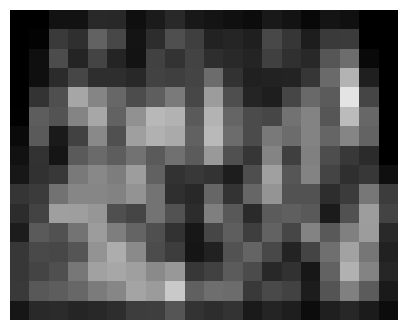}\label{fig:input500_458}} \hfill
\subfloat[PRBPN]{\includegraphics[width=0.16\textwidth]{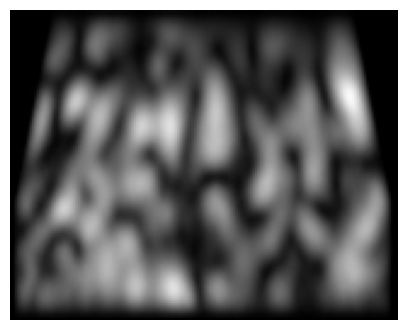}\label{fig:prediction500_458}} \hfill
\subfloat[Fine-mesh PWE]{\includegraphics[width=0.16\textwidth]{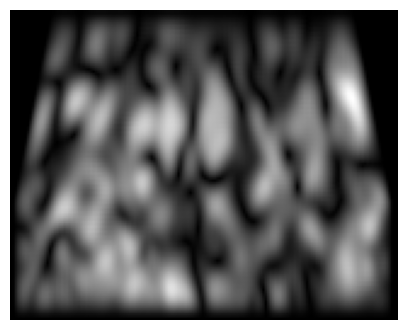}\label{fig:groundtruth500_458}} \hfill
\subfloat[Coarse-mesh PWE]{\includegraphics[width=0.16\textwidth]{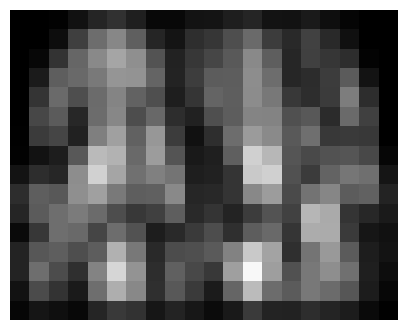}\label{fig:input1000_458}} \hfill
\subfloat[PRBPN]{\includegraphics[width=0.16\textwidth]{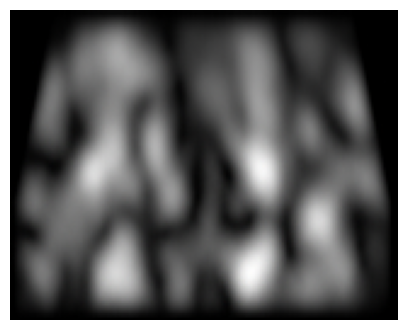}\label{fig:prediction1000_458}} \hfill
\subfloat[Fine-mesh PWE]{\includegraphics[width=0.16\textwidth]{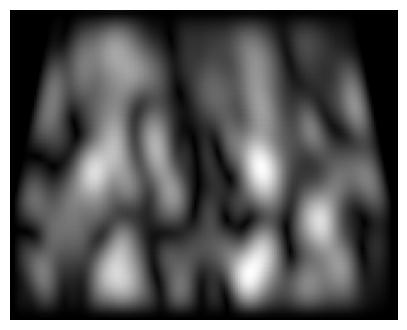}\label{fig:groundtruth1000_458}} \hfill

\caption{Comparison of the results for trapezoidal tunnel under 5.8G: (a) Coarse-mesh PWE at 500m, (b) PRBPN at 500m, (c) Fine-mesh PWE at 500m, (d) Coarse-mesh PWE at 1000m, (e) PRBPN at 1000m, and (f) Fine-mesh PWE at 1000m distances.}
\label{fig:shape458}
\end{figure*}

\subsection{Experimental Indicators}

Model performance is quantified using four standard error metrics that capture complementary aspects of prediction fidelity: mean absolute error (MAE), mean absolute percentage error (MAPE), root mean square error (RMSE), and the coefficient of determination ($R^2$). Their definitions are:

\begin{equation}\label{eq:mae}
    \mathrm{MAE} = \frac{1}{N} \sum_{i=1}^{N} \big| y_i - \hat{y}_i \big|,
\end{equation}

\begin{equation}\label{eq:mape}
    \mathrm{MAPE} = \frac{100\%}{N} \sum_{i=1}^{N} \left| \frac{y_i - \hat{y}_i}{y_i} \right|,
\end{equation}

\begin{equation}\label{eq:rmse}
    \mathrm{RMSE} = \sqrt{ \frac{1}{N} \sum_{i=1}^{N} \big( y_i - \hat{y}_i \big)^2 },
\end{equation}

\begin{equation}\label{eq:r2}
    R^2 = 1 - \frac{ \sum_{i=1}^{N} \big( \hat{y}_i - y_i \big)^2 }{ \sum_{i=1}^{N} \big( y_i - \bar{y} \big)^2 }.
\end{equation}

Here, $y_i$ denotes the reference received power (from Fine-mesh PWE simulations or measurements), $\hat{y}_i$ denotes the prediction form PRBPN, $\bar{y}$ is the mean of $\{y_i\}_{i=1}^N$, and $N$ is the number of sampled spatiotemporal points. Together, \eqref{eq:mae}–\eqref{eq:r2} provide a comprehensive assessment of absolute error, relative error, error dispersion, and explained variance, respectively.

\subsection{Ablation Study on Difference-Weighted Temporal Fusion (DWTF)}
\label{sec:ablation_dwtf}

To demonstrate the necessity and effectiveness of the proposed Difference-Weighted Temporal Fusion (DWTF) module, we conduct an ablation study by removing DWTF while keeping all other components (encoder, iterative projection/back-projection refinement, loss functions, etc.) unchanged. Qualitative comparisons on a representative trapezoidal tunnel at 5.8\,GHz are presented in Fig.~\ref{fig:ablation}.

	
	

As shown in Fig.~\ref{fig:ablation}, the coarse-mesh PWE inputs~\subref{fig:abl_a},~\subref{fig:abl_e} suffer from severe blocky artifacts and blurred multipath patterns due to insufficient spatial resolution. 
The full PRBPN equipped with DWTF~\subref{fig:abl_b},~\subref{fig:abl_f} successfully recovers sharp reflection paths along the slanted walls, fine modal structures, and accurate energy distribution at both 500\,m and 1000\,m, closely matching the fine-mesh PWE references~\subref{fig:abl_d},~\subref{fig:abl_h}. 
In contrast, PRBPN without DWTF~\subref{fig:abl_c},~\subref{fig:abl_g} produces noticeable spurious oscillations, distorted high-intensity regions near the tunnel boundaries, and, more evidently at farther distances, ghosting artifacts together with over-smoothed central energy distribution. 

The superiority of DWTF arises from its physically motivated design: electromagnetic wave propagation in tunnels is a smooth, continuous process governed by the parabolic wave equation, where meaningful spatiotemporal variations are concentrated in regions of strong multipath interaction (wall reflections, mode coupling). By computing pixel-wise differences and generating attention maps directly from these physical changes, DWTF naturally focuses the network on propagation-relevant cues without requiring explicit optical flow estimation or heavy recurrent states. 
It reduces reconstruction artifacts by 10–20\% (see Sec.~II) while also enhancing training stability and data efficiency, providing important benefits in data-scarce, high-frequency tunnel propagation modeling scenarios.


\subsection{Comparison Experiment}

To evaluate the effectiveness of the proposed PRBPN model,  conducted under various tunnel configurations and carrier frequencies. The reference ground truth is provided by high-resolution simulations based on the fine-mesh PWE methods.

The experiments encompass four tunnel cross-section geometries, as illustrated in Fig.~\ref{tunnelshape}. For each geometry, simulations are conducted at four representative carrier frequencies, 0.9\,GHz, 2.4\,GHz, 4.9\,GHz, and 5.8\,GHz, to assess generalization across electromagnetic regimes. At each frequency, four distinct transmitter positions are evaluated to further test robustness. The RSS distribution at selected cross-sections of the tunnel, located 500 m and 1000 m from the initial section, is compared in Figs.~\ref{fig:shape109}-\ref{fig:shape458}. It can be observed that the predictions from PRBPN closely match the results obtained from fine-mesh PWE simulations.


The RSS distribution at selected cross-sections of the tunnel, located 500 m and 1000 m from the initial section, is compared in Figs.~\ref{fig:shape109}-\ref{fig:shape458}. It can be observed that the predictions from PRBPN closely match the results obtained from fine-mesh PWE simulations.

Quantitative evaluations are performed using four standard metrics: MAE, MAPE, RMSE, and \(R^2\). Each dataset consists of 80 tunnel data points, and the results are averaged across all samples. Tables~\ref{shape1_metrics}–\ref{shape4_metrics} present the performance of PRBPN across different tunnel geometries and frequencies. The results demonstrate that PRBPN consistently achieves excellent performance, with \(R^2\) values exceeding 0.85 across all tested configurations and frequencies, and MAE values well within a reasonable range. Notably, PRBPN performs best at lower frequencies, with accuracy consistently above 90\%.


PRBPN achieves exceptional data efficiency, requiring only 10–20 paired coarse/fine samples to reach competitive accuracy, due to its physics-informed design that embeds strong inductive biases. By modeling the longitudinal wave evolution as a short video sequence and performing multi-frame-to-one prediction, the network leverages the inherent temporal smoothness of electromagnetic propagation to guide learning without large datasets. An iterative projection/back-projection refinement enforces low- to high-resolution consistency through error feedback, ensuring energy conservation and propagation continuity while facilitating stable training. In addition, the difference-weighted temporal fusion module aggregates adjacent slices with attention focused on multipath-informative regions, suppressing noise and artifacts while capturing complex interactions efficiently. This combination of physics-based priors and temporal context enables PRBPN to generalize robustly across tunnel geometries and frequencies. Simulations demonstrate that high-fidelity reconstructions can be obtained even with as few as 16 training pairs over long propagation distances.


\begin{table}[ht]
\centering
\caption{Performance of PRBPN under different frequencies for the rectangular tunnel.}
\label{shape1_metrics}
\setlength{\tabcolsep}{3pt}   
\renewcommand{\arraystretch}{1.1}
\begin{tabular}{|c|c|c|c|c|}
\hline
Indicators & 
$f=0.9$~GHz & $f=2.4$~GHz & $f=4.9$~GHz & $f=5.8$~GHz \\
\hline
MAE   & 0.8587 & 1.3578 & 2.7066 & 1.4913 \\
MAPE  & 2.27\% & 3.00\% & 3.60\% & 2.90\% \\
RMSE  & 1.8183 & 3.4699 & 3.2524 & 3.1616 \\
$R^2$ & 0.9753 & 0.9044 & 0.8759 & 0.8950 \\
\hline
\end{tabular}
\end{table}

\begin{table}[ht]
\centering
\caption{Performance of PRBPN under different frequencies for the arched tunnel.}
\label{shape2_metrics}
\setlength{\tabcolsep}{3pt}
\renewcommand{\arraystretch}{1.1}
\begin{tabular}{|c|c|c|c|c|}
\hline
Indicators & $f=0.9$~GHz & $f=2.4$~GHz & $f=4.9$~GHz & $f=5.8$~GHz \\
\hline
MAE   & 0.3103 & 0.5947 & 0.9808 & 1.0676 \\
MAPE  & 1.19\% & 1.74\% & 2.53\% & 2.57\% \\
RMSE  & 0.8053 & 1.6224 & 2.0098 & 2.2177 \\
$R^2$ & 0.9913 & 0.9280 & 0.8843 & 0.8607 \\
\hline
\end{tabular}
\end{table}

\begin{table}[ht]
\centering
\caption{Performance of PRBPN under different frequencies for the arched tunnel with vertical side walls.}
\label{shape3_metrics}
\setlength{\tabcolsep}{3pt}
\renewcommand{\arraystretch}{1.1}
\begin{tabular}{|c|c|c|c|c|}
\hline
Indicators & $f=0.9$~GHz & $f=2.4$~GHz & $f=4.9$~GHz & $f=5.8$~GHz \\
\hline
MAE   & 0.5385 & 0.5472 & 1.0877 & 0.9984 \\
MAPE  & 1.72\% & 1.42\% & 2.37\% & 2.17\% \\
RMSE  & 1.1431 & 1.4623 & 2.2083 & 2.1878 \\
$R^2$ & 0.9922 & 0.9488 & 0.8586 & 0.8669 \\
\hline
\end{tabular}
\end{table}

\begin{table}[ht]
\centering
\caption{Performance of PRBPN under different frequencies for the trapezoidal tunnel.}
\label{shape4_metrics}
\setlength{\tabcolsep}{3pt}
\renewcommand{\arraystretch}{1.1}
\begin{tabular}{|c|c|c|c|c|}
\hline
Indicators & $f=0.9$~GHz & $f=2.4$~GHz & $f=4.9$~GHz & $f=5.8$~GHz \\
\hline
MAE   & 0.7304 & 0.8067 & 1.2313 & 1.3241 \\
MAPE  & 2.40\% & 2.12\% & 2.72\% & 2.74\% \\
RMSE  & 1.3641 & 1.7921 & 2.3434 & 2.5036 \\
$R^2$ & 0.9833 & 0.9414 & 0.8672 & 0.8505 \\
\hline
\end{tabular}
\end{table}

\begin{table}[ht]
\centering
\caption{Setup of the input parameters for the propagation simulations in Massif Central tunnel.}
\label{TableII}
\setlength{\tabcolsep}{3pt}
\renewcommand{\arraystretch}{1.1}
\begin{tabular}{|c|c|c|c|}
\hline
Parameter & Min.\ value & Max.\ value & Increment \\
\hline
$f$ [GHz]          & 0.9    & 2.1   & 1.2   \\
$x_{\text{TX}}$ [m] & 0      & 2.0   & 0.5   \\
$y_{\text{TX}}$ [m] & 0.5    & 3.0   & 0.5   \\
$z$ [m]            & 0      & 2500  & 0.25  \\
$x_{\text{RX}}$ [m] & $-1.5$ & 1.5   & 0.15  \\
$y_{\text{RX}}$ [m] & 0.2    & 3.2   & 0.15  \\
\hline
\end{tabular}
\end{table}

\section{Application: Massif Central Tunnel}\label{Section IV}

\begin{figure}[H]
	\centering
	\includegraphics[width=0.2\textwidth]{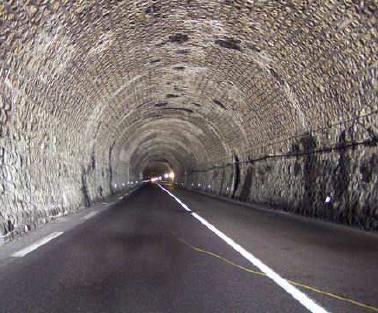}
	\includegraphics[width=0.2\textwidth]{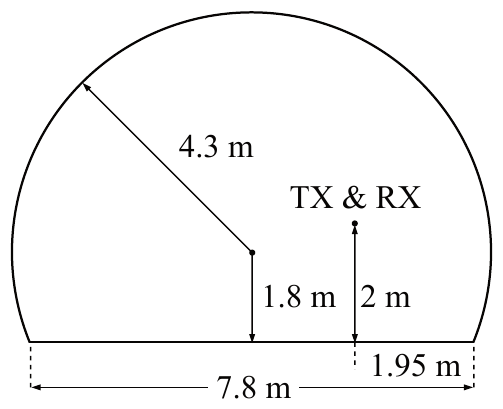}
	
	\caption{Cross-section geometry and a on-site image of the Massif Central tunnel in France.}
	\label{massif_photo}
\end{figure}

To further substantiate the industrial applicability and robustness of the proposed PRBPN framework, we conduct an engineering validation using empirical measurements collected from the Massif Central tunnel in France~\cite{Dudley07}. This tunnel is a long, enclosed environment primarily constructed from large stone blocks and reinforced concrete, as shown in Fig.~\ref{massif_photo}.

The tunnel boundary materials are modeled with relative permittivity \(\varepsilon_r = 5\) and electrical conductivity \(\sigma = 0.01\,\mathrm{S/m}\). Two operating frequencies are considered: \(0.9\,\mathrm{GHz}\) and \(2.1\,\mathrm{GHz}\).
The simulation setup used during training is summarized in Table~\ref{TableII}. Given the relatively low operating frequencies and the correspondingly smoother spatial variation of the received signal compared with the synthetic cases in Section~\ref{Section III}, we adopt a larger spatial scale factor for efficiency. High-resolution PWE simulations use \(\Delta x=\Delta y=0.4\lambda\) and \(\Delta z=2\lambda\), as previously described. For the coarse-mesh configuration, the discretization is \(\Delta x=\Delta y=3.2\lambda\) and \(\Delta z=2\lambda\), maintaining consistency with prior configurations.

\begin{figure}[ht]
	\centering
	\includegraphics[width=0.4\textwidth]{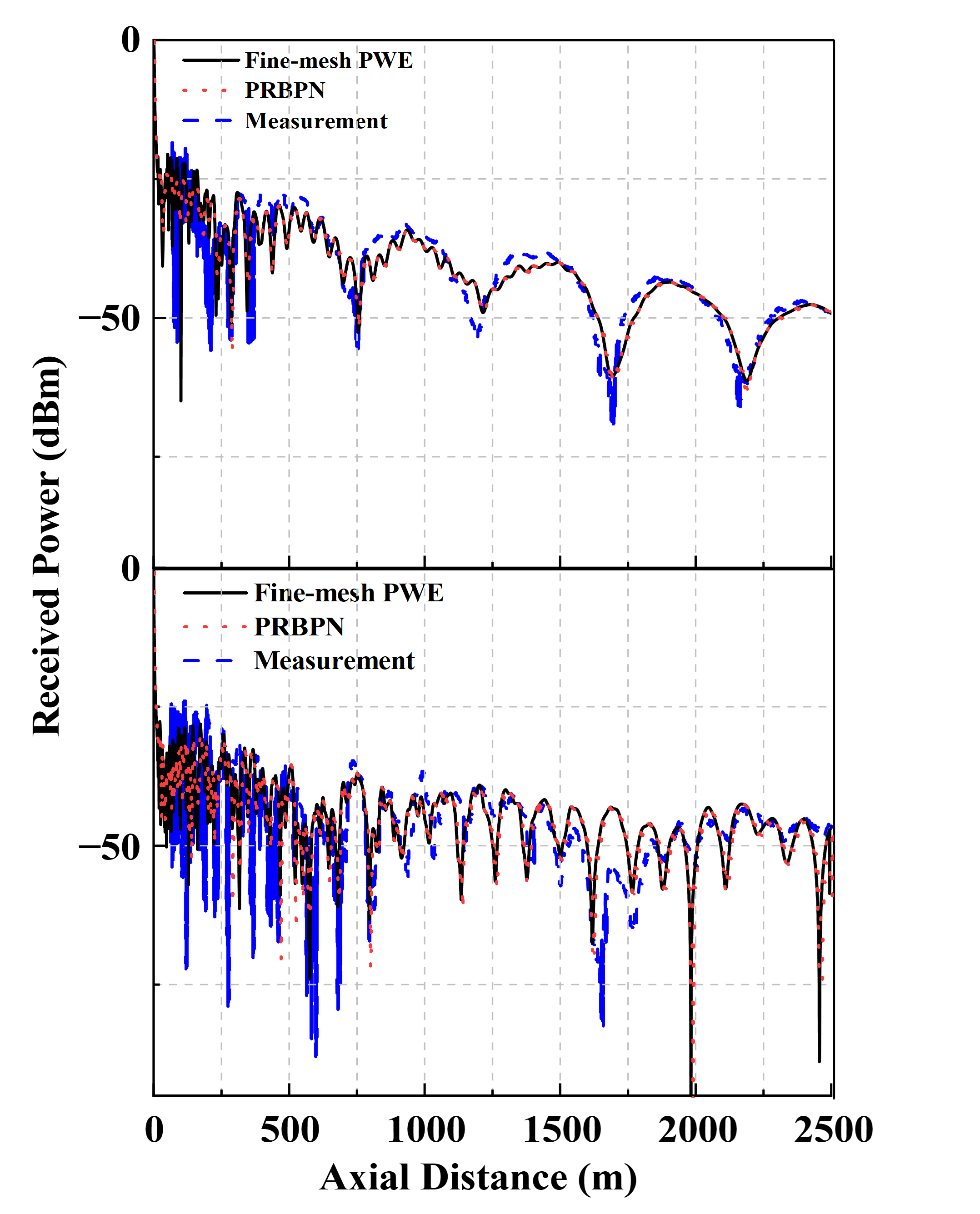}
	\caption{Comparison of received power distributions along the Massif Central tunnel at 0.9 GHz and 2.1 GHz.}
	\label{massif_prediction}
\end{figure}

Prediction performance is illustrated in Fig.~\ref{massif_prediction}, where PRBPN’s received-power distributions are compared against fine-mesh PWE references. Consistent with the synthetic validations, PRBPN exhibits close agreement with ground truth and accurately captures long-range propagation characteristics.
Quantitatively, Table~\ref{tab:meas_metrics} shows that PRBPN attains high-fidelity performance across both frequencies in this real-world setting. At \(0.9\,\mathrm{GHz}\), the model achieves \(\mathrm{MAE}=0.4192\), \(\mathrm{MAPE}=1.32\%\), \(\mathrm{RMSE}=1.2544\), and \(R^2=0.9780\), indicating near-perfect alignment with the fine-mesh PWE reference. At \(2.1\,\mathrm{GHz}\), where multipath and attenuation are more pronounced due to shorter wavelength, PRBPN maintains strong accuracy with \(\mathrm{MAE}=0.4947\), \(\mathrm{MAPE}=1.22\%\), \(\mathrm{RMSE}=1.4903\), and \(R^2=0.9520\). These results reflect the model’s capacity to capture propagation dynamics and adapt to varying electromagnetic conditions.

\begin{figure}[ht]
	\centering
	\subfloat[Coarse-mesh PWE]{\includegraphics[width=0.16\textwidth]{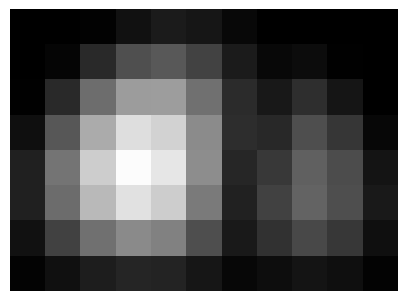}\label{fig:input500_458}} \hfill
	\subfloat[PRBPN]{\includegraphics[width=0.16\textwidth]{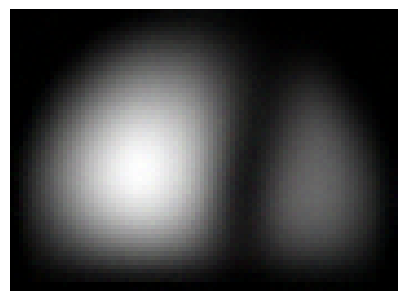}\label{fig:prediction500_458}} \hfill
	\subfloat[Fine-mesh PWE]{\includegraphics[width=0.16\textwidth]{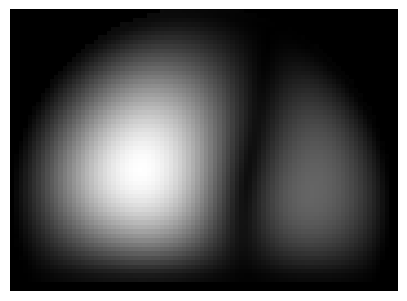}\label{fig:groundtruth500_458}} \hfill
	\subfloat[Coarse-mesh PWE]{\includegraphics[width=0.16\textwidth]{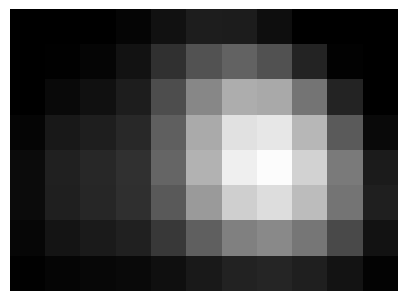}\label{fig:input1000_458}} \hfill
	\subfloat[PRBPN]{\includegraphics[width=0.16\textwidth]{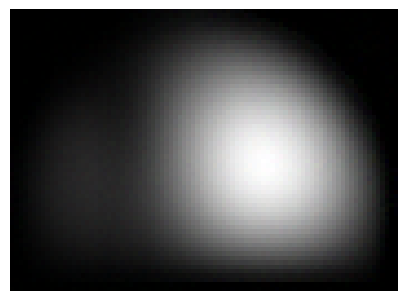}\label{fig:prediction1000_458}} \hfill
	\subfloat[Fine-mesh PWE]{\includegraphics[width=0.16\textwidth]{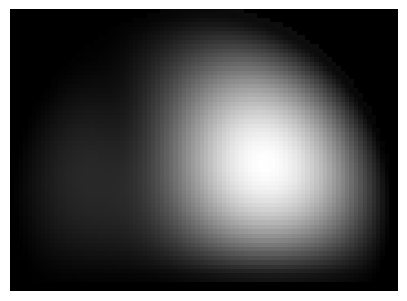}\label{fig:groundtruth1000_458}} \hfill
	
	\caption{Comparison of the results for Massif Central tunnel under 0.9G: (a) Coarse-mesh PWE at 1250m, (b) PRBPN at 1250m, (c) Fine-mesh PWE at 1250m, (d) Coarse-mesh PWE at 2500m, (e) PRBPN at 2500m, and (f) Fine-mesh PWE at 2500m distances.}
	\label{realshape209}
\end{figure}

\begin{figure}[ht]
	\centering
	\subfloat[Coarse-mesh PWE]{\includegraphics[width=0.16\textwidth]{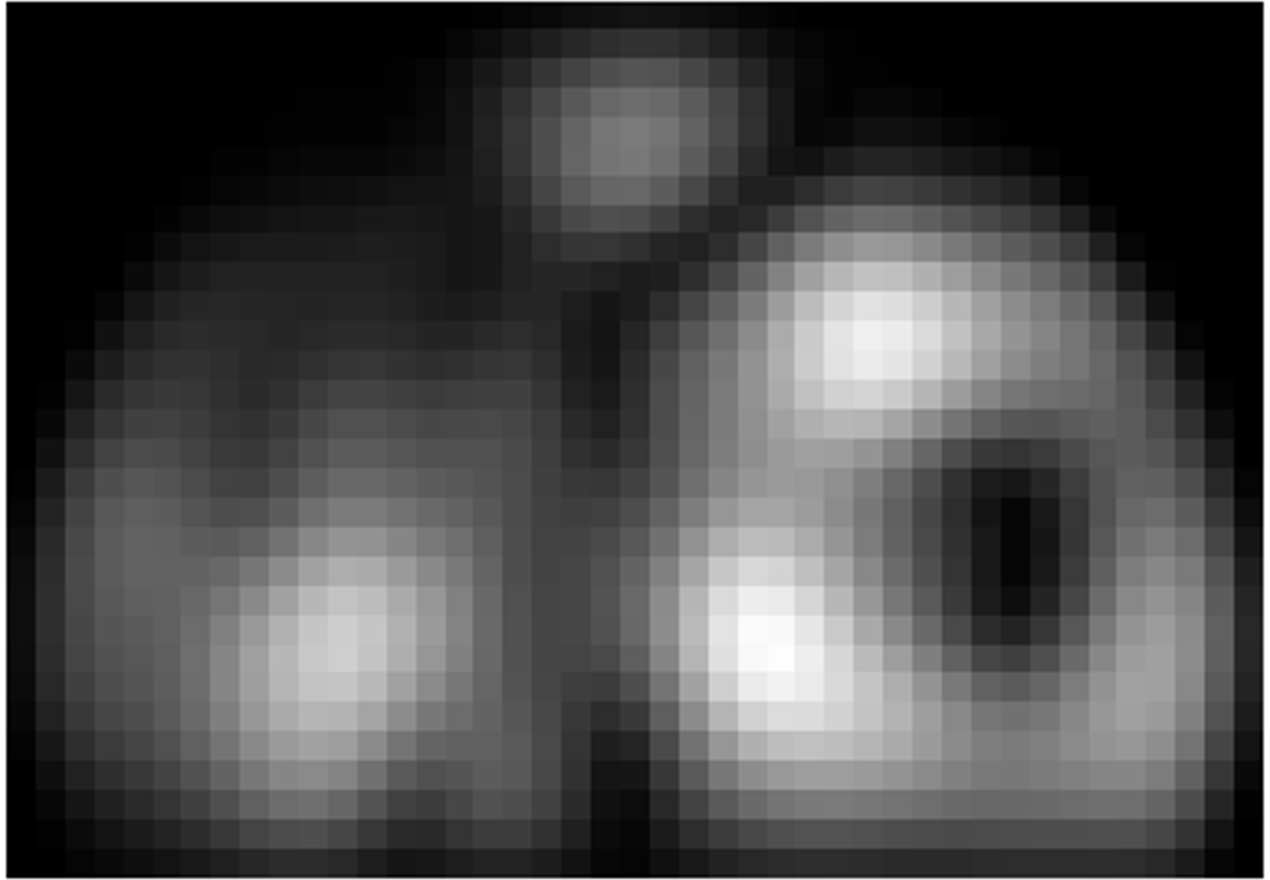}\label{fig:input500_458}} \hfill
	\subfloat[PRBPN]{\includegraphics[width=0.16\textwidth]{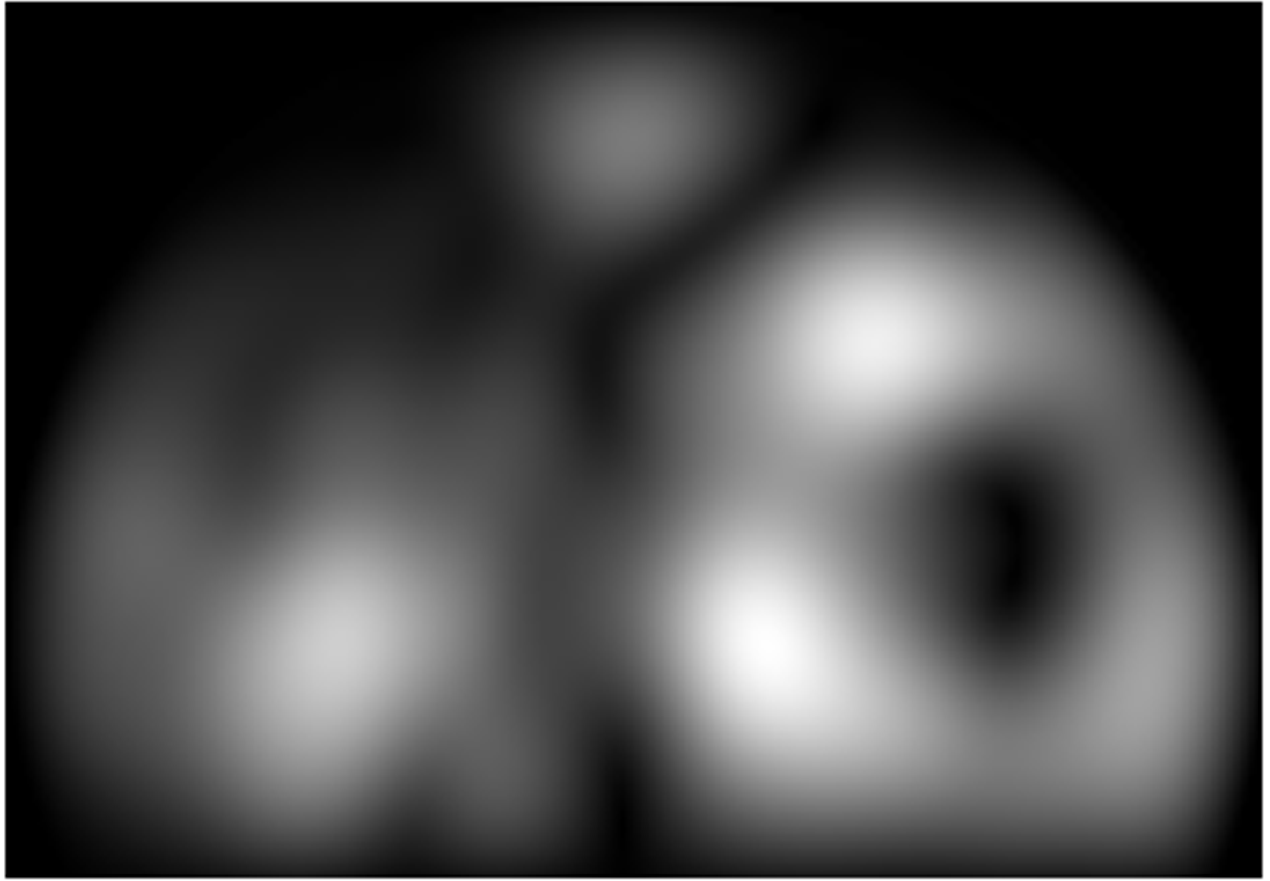}\label{fig:prediction500_458}} \hfill
	\subfloat[Fine-mesh PWE]{\includegraphics[width=0.16\textwidth]{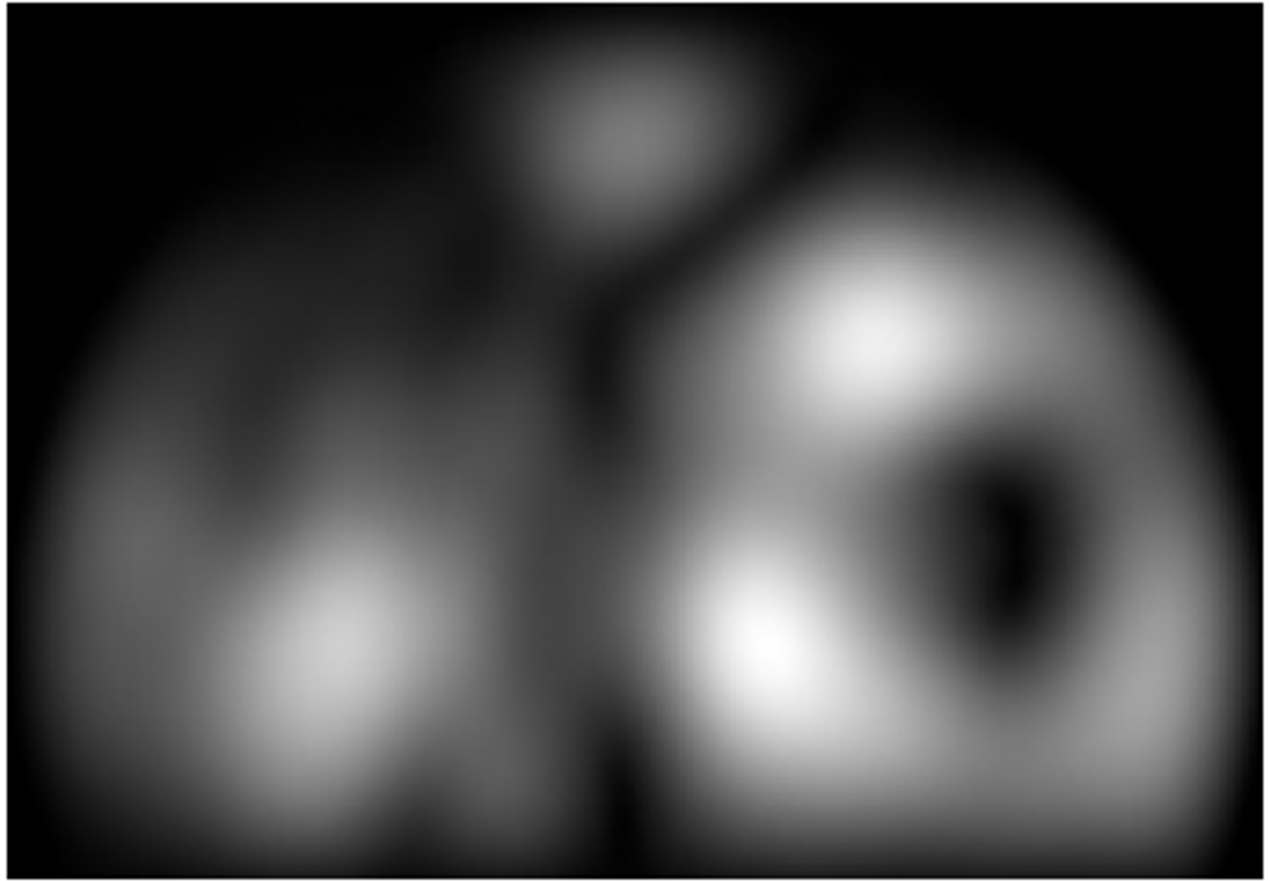}\label{fig:groundtruth500_458}} \hfill
	\subfloat[Coarse-mesh PWE]{\includegraphics[width=0.16\textwidth]{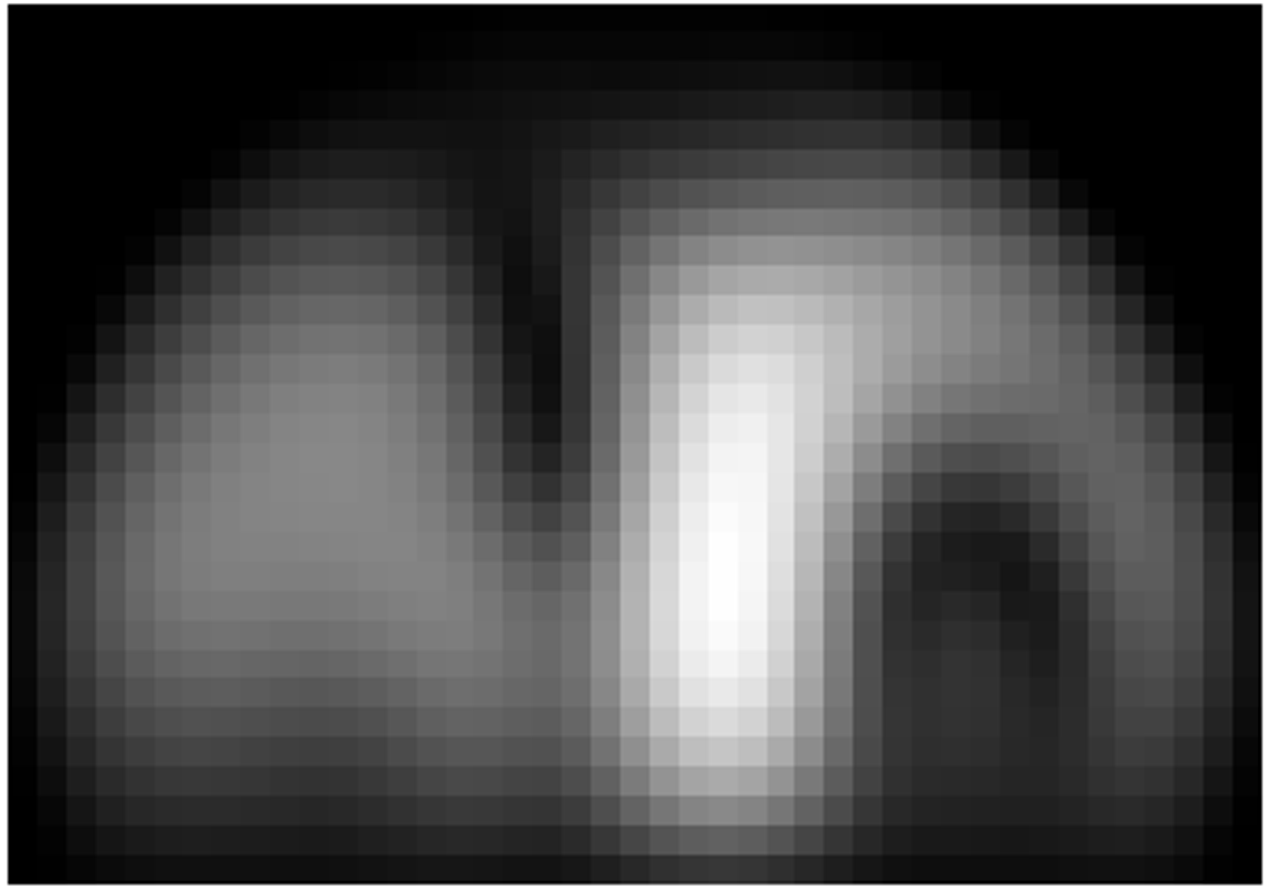}\label{fig:input1000_458}} \hfill
	\subfloat[PRBPN]{\includegraphics[width=0.16\textwidth]{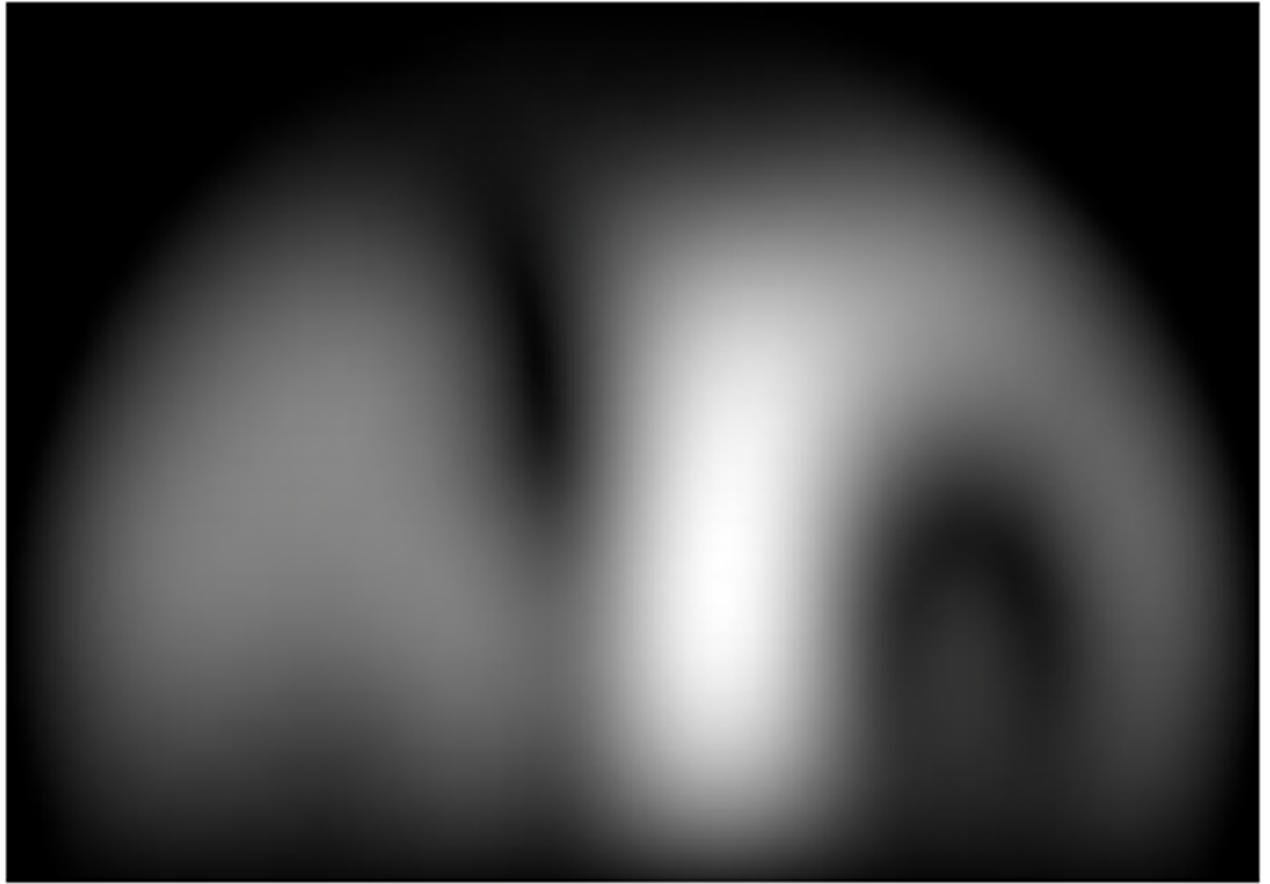}\label{fig:prediction1000_458}} \hfill
	\subfloat[Fine-mesh PWE]{\includegraphics[width=0.16\textwidth]{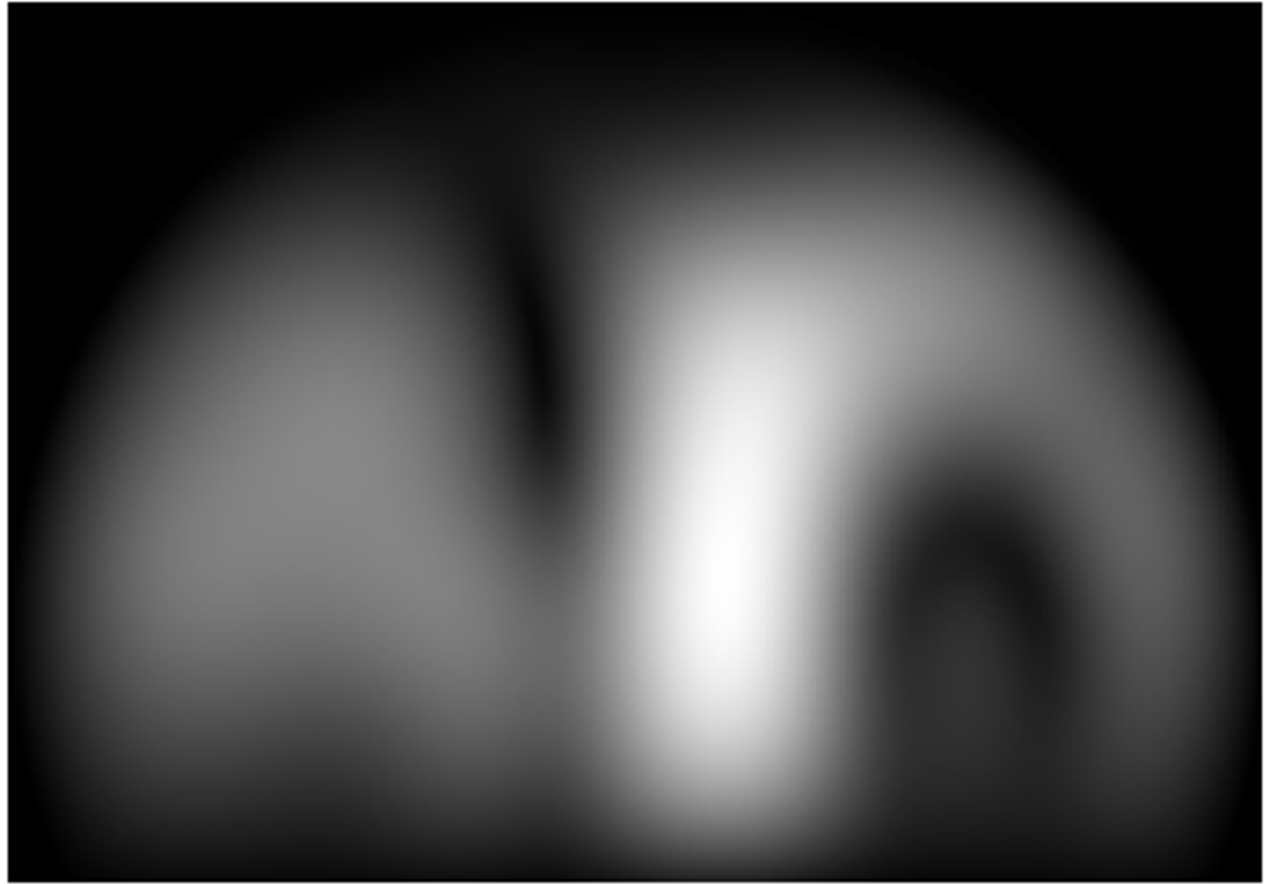}\label{fig:groundtruth1000_458}} \hfill
	
	\caption{Comparison of the results for Massif Central tunnel under 2.1G: (a) Coarse-mesh PWE at 1250m, (b) PRBPN at 1250m, (c) Fine-mesh PWE at 1250m, (d) Coarse-mesh PWE at 2500m, (e) PRBPN at 2500m, and (f) Fine-mesh PWE at 2500m distances.}
	\label{realshape221}
\end{figure}

Detailed cross-sectional comparisons are provided in Figs.~\ref{realshape209} and~\ref{realshape221}. At \(1250\,\mathrm{m}\) and \(2500\,\mathrm{m}\), PRBPN reconstructions (subfigures (b) and (e)) closely reproduce the energy concentration zones, wall-reflection paths, and higher-order modal patterns observed in the fine-mesh PWE results (subfigures (c) and (f)), whereas the coarse-mesh inputs (subfigures (a) and (d)) are visibly blurred and contaminated by numerical artifacts. Notably, at the far-end \(2500\,\mathrm{m}\) section, PRBPN recovers fine-grained textures and attenuation gradients with errors substantially below those of the coarse-grid simulations, underscoring robust long-range fidelity in a complex real-world tunnel.

Importantly, these results are obtained under a data-scarce regime: for each frequency, only 4 paired coarse/fine tunnels are used for training. The observed performance gains are enabled by the DWTF module’s attention-weighted fusion of multipath-informative regions together with the iterative projection/back-projection mechanism, which enforces physics-consistent error correction without reliance on large training corpora.

\begin{table}[ht]
	\centering
	\caption{Performance of PRBPN under both frequencies for Massif Central tunnel.}
	\label{tab:meas_metrics}
	\setlength{\tabcolsep}{3pt}
	\renewcommand{\arraystretch}{1.1}
	\begin{tabular}{|c|c|c|}
		\hline
		Indicators & $f=0.9$~GHz & $f=2.1$~GHz \\
		\hline
		MAE   & 0.4192 & 0.4947 \\
		MAPE  & 1.32\% & 1.22\% \\
		RMSE  & 1.2544 & 1.4903 \\
		$R^2$ & 0.9780 & 0.9520 \\
		\hline
	\end{tabular}
\end{table}

\section{Conclusion}
\label{sec:conclusion}

This work introduced PRBPN, a physics-informed recurrent back-projection framework for fast and accurate tunnel propagation modeling from coarse-grid PWE inputs. Leveraging difference-weighted temporal fusion and iterative projection/back-projection for LR-to–HR consistency, PRBPN reconstructs fine-grid RSS fields without a pre-upsampling stage and with strong data efficiency. Across four representative tunnel cross-sections and four carrier frequencies, PRBPN closely matches fine-mesh PWE references, achieving consistently low MAE/MAPE/RMSE and high $R^2$. Real-world validation on France’s Massif Central tunnel further confirms robustness: trained with only 4 paired coarse/fine samples per frequency, PRBPN maintains high fidelity at both 0.9 and 2.1\,GHz.


\vfill\pagebreak

\end{document}